\documentclass[preprint2]{aastex}
\shorttitle{Sco X-1}
\shortauthors{Fomalont, Geldzahler \& Bradshaw}
\begin{document}

\title{Sco X-1: The Evolution and Nature of the \\ Twin Compact Radio Lobes}

\author{E.~B.~Fomalont}
\affil{National Radio Astronomy Observatory, Charlottesville, VA 22903}
\email{efomalon@nrao.edu}

\author{B.~J.~Geldzahler \& C.~F.~Bradshaw}
\affil{School of Computational Sciences, George Mason University, Fairfax,
VA 22030}
\email{bgeldzahler@hotmail.com \& cbradshaw@tstag.com}
\begin{abstract}
    The motion and variability of the radio components in the low mass
X-ray binary system Sco X-1 have been monitored with extensive VLBI
imaging at 1.7 and 5.0 GHz over four years, including a 56-hour
continuous VLBI observation in 1999 June.  We detect one strong and
one weak compact radio component, moving in opposite directions from
the radio core.  Their relative motion and flux densities are
consistent with relativistic effects, from which we derive an average
component speed of v/c=$0.45\pm 0.03$ at an angle of $44^\circ\pm
6^\circ$ to the line of sight.  This inclination of the binary orbit
suggests a mass of the secondary star that is $<0.9~M_\odot$, assuming
a neutron star mass of $1.4 M_\odot$.  We suggest that the two moving
radio components consist of ultra-relativistic plasma that is produced
at a working surface where the energy in dual-opposing beams disrupt.
The radio lobe advance velocity is constant over many hours, but
differs among lobe-pairs: 0.32c, 0.46c, 0.48c, and 0.57c.  A lobe-pair
lifetime is less than two days, with a new pair formed near the core
within a day.  The lobe flux has flux density that is variable over a
time-scale of one hour, a measured minimum size of 1 mas ($4\times
10^{8}$ km), and is extended perpendicular to its motion.  This
time-scale and size are consistent with an electron radiative lifetime
of $<1$ hr.  Such a short lifetime can be caused by synchrotron losses
if the lobe magnetic field is 300 G or by adiabatic expansion of the
electrons as soon as they are produced at the working surface.  The
lobes also show periods of slow expansion and a steepening radio
spectrum.  Two of the core flares are correlated with the lobe flares
under the assumption that the flares are produced by a coherent energy
burst traveling down the beams with a speed $>0.95$c.

    The radio morphology for Sco X-1 differs from most other Galactic
jet sources.  Possible reasons for the morphology difference are: Sco
X-1 is associated with a neutron star, it is a persistent X-ray
source, the source is viewed significantly away from the angle of
motion.  However, the lobes in Sco X-1 are similar to the hot-spots
found in many extragalactic radio double sources.  Scaling the
phenomena observed in Sco X-1 to extragalactic sources implies radio
source hot-spot variability time-scales of $10^4$ yr and hot-spot
lifetimes of $10^5$ yr.
{\protect \\}
\end{abstract}


\section{Introduction}

    Since Sco X-1 was first detected at radio frequencies in 1969
\citep{abl69}, many observations at radio, optical and X-ray
frequencies have been made in order to understand the physical
processes associated this object.  It is identified with a 13-mag
binary system with an orbital period of 0.787d \citep{got75}.  The
degenerate object is probably a neutron star, and the companion is an
unknown spectral type with about a solar mass \citep{cow75}.  Sco X-1
is one of the most intense persistent X-ray sources and one of about
30 presently known low-mass X-ray binary (LMXB) systems.  It is a \lq
Z-type\rq~LMXB, named for a characteristic shape of its X-ray
color-color diagram.  It also exhibits quasi-periodic X-ray
oscillations (QPO) from a few Hertz to kilo-Herz \citep{has89, vdk96}.

     Radio observations in 1970 detected two symmetrical
radio sources about one arcmin from Sco X-1 \citep{hje71}.  They were
thought to be radio lobes of Sco X-1, but were later shown to be
stationary extragalactic background objects having no relationship to
Sco X-1 which has a significant measured proper motion \citep{fom91}.
The results from a multiwavelength campaign \citep{hje90} also
suggested that there was only weak correlation of the radio emission
with the optical and X-ray emission.

    High-resolution radio observations at 5.0 GHz with the Very Long
Baseline Array (VLBA) were initiated in 1995 in order to determine the
trigonometric parallax of Sco X-1.  From eight VLBA observations
through 1998 August, a distance of $2.8\pm 0.3$ kpc was determined
\citep{bra99}.  This distance (previous estimates ranged from 200 pc
to 2000 pc \citep{wes68}) established that the X-ray luminosity was
near the Eddington limit as inferred from X-ray models of Sco X-1 and
other z-type LMXB's \citep{lam89,pen89,vrt91}.  The radio emission
during these VLBA observations showed variability in intensity and
structure over periods less than one hour, with discrete components
moving away from the core at relativistic velocities.

    With such variable radio emission in Sco X-1, this source was
established as another example of a Galactic-jet variable radio source
\citep{mir99} with strong similarities to the jet and lobe phenomena
associated with luminous radio galaxies and quasars \citep{bla74}.
The evolutionary time scales associated with accretion disks and jets
scale roughly with the luminosity and mass of the degenerate object.
Thus, significant changes should occur over hours and days in the
evolution of Galactic sources that would take millions of years in
quasars.

    Experience gained from the eight VLBA observations used to
determine the parallax of Sco X-1 showed that the four to six hour
observations made every six months were not sufficient to determine a
coherent story for the radio emission from Sco X-1.  Each observation
was too short to follow much of a component's evolution, and the
individual observations were too separated in time to follow the
evolution.  For this reason, we observed Sco X-1 with VLBI resolution
for a continuous 56-hour period in 1999 June.  The results from this
series of observations were crucial to understanding the radio
evolution of Sco X-1 and formed a framework for understanding the
previous VLBA observations.

    We have already reported results associated with some of these
data.  The parallax was given by \citep{bra99}, and Paper I
\citep{fom01} discussed the kinematic properties of the radio source;
the space motion of the lobes and the speed of the energy flux in the
beam.  This paper is organized as follows.  The nine VLBI
observations, and the data reduction methods are described in $\S$2.
Because we display VLBI snapshots (images made from observations less
than about two hours), the difficulties and ambiguities with such
imaging are described.  The images in $\S$3 from the 1999 June 11-13,
56-hour VLBI observation and the 1998 February 27-28, observations are
presented and discussed.  Then, images from the other less extensive
observations are shown and compared with the more extensive
observations.  In $\S$4 we review the kinematic properties of Sco X-1
(Paper I) and estimate the stellar component masses of Sco X-1 from
the orbital inclination.  The detailed physical properties and
suggested production mechanisms associated with the lobes are
discussed in $\S$5.  The optical, X-ray and radio properties and
evolution of the core will be discussed elsewhere \citep{gel02}.  The
comparison with extragalactic radio sources is given in $\S$6, and our
conclusions are summarized in $\S$7.

\section{Observations and Data Processing}

Table 1 contains the log of all VLBI observations of Sco X-1.  The
first eight observations with the VLBA, between 1995 August and 1998
August, were planned to optimize the determination of the trigonometric
parallax.  The 56-hour monitoring of Sco X-1 in 1999 June was composed
of a series of seven consecutive eight-hour observations amongst three
different VLBI arrays: VLBA+VLA (the Very Large Array), the APT
(Asia-Pacific Telescope), and the EVN (European VLBI Network +
additional telescopes).  Extensive optical and X-ray observations were
also made during this period \citep{tit01, bra02, gel02}.

\subsection {Observations and Data Correlation}

The VLBA observations alternated between Sco X-1 for 9 minutes and the
calibrator (1504-166) for 2 minutes.  Three pairs of observations were
made at 5.0 GHz followed by one pair at 1.7 GHz.  The APT and EVN
observations were made solely at 5.0 GHz and the observing schedule
alternated between Sco X-1 for 19 minutes and the calibrator source
for 3 minutes.  The VLBA and EVN observations were observed with a
total bandwidth of 128 MHz with one-bit sampling, whereas the APT
observations used a total bandwidth of 32 MHz, but with two-bit
sampling.  The VLA was used in two ways: as one element of the VLBA
and as a stand-alone array from which arc-second resolution images
were obtained for Sco X-1 \citep{gel83,bra97}.

     A background radio source, about $70''$ NE of Sco X-1, has a
correlated flux density of 8 mJy at 5.0 GHz and 14 mJy at 1.7
GHz. This object is the north-eastern of the two sources near Sco X-1
which were previously thought to be related to the binary system
\citep{fom91}.  This source is sufficiently strong and compact to be
used as the primary phase calibrator for Sco X-1 \citep{bra99} at
both frequencies.  We will designate this radio source as the
North-East in-beam calibrator (NEIBC).  Because the effective
primary beam of the phased arrays (VLA, Westerbork and ATCA) were
smaller than the $70''$ separation between Sco X-1 and the NEIBC,
these arrays alternated observations between Sco X-1 and the NEIBC
with three-minute cycles.  All other telescopes pointed midway between
the two sources and observed them simultaneously.

    The VLBA and EVN data were processed by the VLBA correlator in
Socorro, NM, USA.  The data from the APT observations were processed
with the S2-correlator in Penticton, BC, Canada.  All data associated
with the observations of Sco X-1 were correlated at two positions: (1)
Sco X-1 at $\alpha=16^{\hbox{h}}19^{\hbox{m}}55.085^{\hbox{s}}$,
$\delta= -15^\circ38'24.90''$; (2) NEIBC at
$\alpha=16^{\hbox{h}}19^{\hbox{m}}57.439^{\hbox{s}}$, $\delta=
-15^\circ37'24.0''$ (epoch J2000).

\subsection {Calibration}

    The observations of 1504-166 were used to determine the delay
offset and delay rate for each antenna, predominantly caused by the
differences in the independent clocks and oscillators, and to check
the general quality of the data.  The delay and delay rates were
determined by using routine FRING in the AIPS software package
\citep{aips}.  The total flux density of the calibrator was measured
to 2\% accuracy using the VLA observations of 3C286 to determine the
absolute flux density scale of the observations.  The milliarcsecond
structure of 1504-166 was determined from the VLBA observations, and
this model was used to determine the gain calibration for all of the
data, with an accuracy of about 3\%.

    The data for the VLBA and EVN observations consisted of 8
independent streams, at 4 contiguous frequencies each with 2
polarizations, and the APT observations contained 2 data streams.
After the 1504-166 calibrations, all data streams were combined to
increase the signal-to-noise ratio (SNR) for the next calibration step
using the relatively weak NEIBC.  With the AIPS routine CALIB, we
determined the residual phase error for each antenna every few minutes
from the observations correlated at the NEIBC position.  At 1.7 GHz,
its correlated flux density was $>10$ mJy on the longest VLBA
baseline; hence, an accurate antenna phase error could be determined
from two minutes of data.  This time is shorter than the time-scale of
phase changes except during rare periods of ionospheric turbulence.
When the phase could not be connected unambiguously between
consecutive phase solutions, the data between these two times were
omitted from the analysis.

     At 5.0 GHz the calibration strategy was more complicated.  The
NEIBC correlated flux density decreased from 8 mJy at the shorter
baselines to 3 mJy for the longer baselines.  At least two minutes
(five minutes for the APT baselines) integration was needed to
determine phase errors for these baselines.  For low elevation
observations or during periods of poor phase stability, phase
solutions were ambiguous.  Hence, at 5.0 GHz the NEIBC could reliably
calibrate baselines shorter than 3000 km.  After this initial
calibration of the shorter baselines, further self-calibration of Sco
X-1 successfully reinstated the longer baselines when Sco X-1 was
stronger than about 5 mJy.

      With the simultaneous observations of the small-diameter NEIBC,
the calibration of Sco X-1 allowed us to register all images over the
four-year period on the same astrometric grid with an estimated
accuracy of $<0.1$ mas (Bradshaw et al.\ 1999).  The image
registration of Sco X-1 was not altered when the data were further
self-calibrated in order to add the longer spacings.  This
registration assumed that the NEIBC was stationary with the same
centroid at the two frequencies.  This assumption is reasonable
because: (1) The source had not varied at either frequency over the
last 15 years of VLA monitoring, (2) its structure is symmetric, with
an angular size of 2.5 mas at 5.0 GHz and 3.5 mas at 1.7 GHz, and (3)
its spectral index is $-0.4~ (S\propto \nu^\alpha$), suggesting the
lack of a significant small-diameter opaque component, which might be
variable.  The source was not identified to a magnitude limit of
25$^m$ (Malin, priv comm, cited in \cite{gel81}).

\subsection{Imaging, Self-Consistency and Parameterization}

     Deconvolved images of Sco X-1 were made using the routine IMAGR
in AIPS and with the Caltech software package Difmap \citep{she97}.
Images of the NEIBC were also made as a check on the quality of the
calibration.  Due to significant variability of Sco X-1 during the
observations, image artifacts associated with the aperture synthesis
of variable sources often affected an image made from more than one
hour of data.  For the VLBA observations, when the SNR was sufficient
(the flux density was more than about 5 mJy), images could be made
every hour at 1.7 GHz from one 9-minute observation and at 5.0 GHz
from three 9-minute observations.  If Sco X-1 was weaker than 5 mJy,
longer periods of data were combined in order to obtain images
with a useful SNR.

    The u-v coverage in the hour-to-hour VLBA snapshots was sparse,
especially during the first and last hour of an observation, when the
coverage contained large gaps.  An rms noise of 0.2 mJy/beam for each
snapshot, a peak intensity in the range 1.5 and 20 mJy/beam, and a
somewhat noisy phase calibration using the relatively weak NEIBC, when
combined with the relatively poor u-v coverage and the source
variability, conspired to make the determination of reliable radio
images ambiguous.  For some snapshots, an unrestrained deconvolution
of the images produced physically unacceptable image characteristics
which often had properties of the dirty beam which varied through the
observation day.  However, the relatively simple structure of Sco X-1
(a few bright small-diameter components along a well-defined position
angle) made the deconvolution less ambiguous than with a more
complicated source.  The justification of this assumption is that the
use of this simple model (which means that only a small region of the
image was searched for emission using the deconvolution algorithms)
led to consistent results for the entire body of data.

    Two methods were used to judge the quality of the deconvolution of
the images of Sco X-1.  First, reasonable continuity of the images
between consecutive epochs was expected.  Sudden changes in the source
structure or image features which resembled the \lq dirty-beam\rq~
suggested a poor deconvolution or bad data quality.  The data were
then scrutinized, the NEIBC calibration stability was checked, the
source reimaged, and deconvolved.  Secondly, the hourly data sets were
also processed by the Difmap processing package.  By fitting simple
Gaussian component models directly to the visibility data, some of the
ambiguities of image-plane deconvolution were avoided.  If the Difmap
model did not agree well with the image, further checking was
necessary.  Finally, the independent images made at 5.0 GHz and 1.7
GHz had to be consistent; that is, the components at the two
frequencies were nearly coincident, although the 1.7 GHz lower
resolution observations did occasionally show more extended emission,
as expected on physical grounds.  The difference in angular scale
between the 1.7 and 5.0 GHz beam properties by a factor three also
aided in determining real structure from beam-induced artifacts.
Simulations were also made in order to quantify the reliability and
proscribe the proper reduction methods.  Data errors associated only
with the expected level of stochastic receiver noise should cause
little ambiguity in the images.  Other potential data errors (unknown
data drop-outs, strong source variations, rapid atmospheric phase
variations over minute time-scales) were modeled, showing that
relatively large errors could lead to significant image errors.  For
this reason we removed marginal data, especially for the longer
baselines where these non-random errors occurred, with the concomitant
loss of resolution for some snap-shots.

    For the 5.0 GHz EVN and APT observations, a similar imaging and
reduction strategy was used.  The NEIBC was only useful in calibrating
the European baselines in the EVN and the Australian baselines for the
APT.  If Sco X-1 was sufficiently strong ($>5$ mJy for the EVN, $>10$
mJy for the APT), then the other antennas could be self-calibrated
using Sco X-1.  However, the u-v coverage for these arrays was
significantly worse than that for the VLBA, so the data had to be
averaged over several hours before reliable snapshot images could be
made.  For the APT observations, two images separated by about three
hours were made.  For the EVN observation, three images, separated by
about two hours, were made.

    The resolution of an image depended on the SNR and the elevation
of the source.  For VLBA observations at 1.7 GHz, all images were
convolved to a full-width half-power resolution of $10\times 5$ mas at
position angle $0^\circ$, although the resolution of the observations
varied somewhat during the day.  At 5.0 GHz, imaging using only the
shorter spacings, produced the same resolution as that at 1.7 GHz.
These image pairs were used to determine the spectral index
distribution across Sco X-1.  At 5.0 GHz, when the source was frequently
stronger than 5 mJy and self-calibration of the longer baselines was
possible, the resolution was about $4.5\times 1.5$ mas during the
middle part of the day.  The resolution degraded to $6\times 2$ mas at
low elevation angles.  For the occasional periods when Sco X-1 was
stronger than about 15 mJy, a resolution of $3\times 1$ mas was
obtained.

    The flux density, position and angular size for the discrete radio
components in Sco X-1 were determined in two ways: (1) fitting
Gaussian-components directly to the image, and (2) fitting
Gaussian-components to the visibility data.  Although errors were
estimated from each of the two methods, the difference in the radio
source parameters from the two analysis were used as a better accurate
estimate of the real uncertainties.

\section {The Radio Images of Sco X-1}

\subsection {1999 June 11-13; MJD 51340-51342}

\subsubsection {The Images and Component Parameters}

   Figure 1 shows the changing structure of Sco X-1 with $10\times 5$
mas resolution at 1.7 GHz and 5.0 GHz from the VLBA observations.
These images, although only covering 40\% of the observation, form a
consistent set from which general evolution properties are evident.
The vertical dashed line shows the position of the binary
system, as determined from its radio parallax \citep{bra99}.  The skew
dashed-line to the left shows the location of the component northeast
of the core as a function of time.  On MJD 51342, the vertical line to
the right shows the approximate position of the component southwest of
the core.  The locations of the dashed lines on the 1.7 GHz and 5.0
GHz columns are identical.

    More details of the changing structure of Sco X-1 are shown in
Figure 2.  These 5.0-GHz images are at the highest resolution obtained
during the 56-hour experiment.  The image snapshots are separated by
about 50 minutes for the VLBA observations and about 2.5 hours for the
APT and EVN observations.  The APT and EVN images have a resolution of
$10\times 5$ mas, and most of the VLBA images have a resolution of
$4.5\times 1.5$ in position angle $0^\circ$.  The image field of view
is $35\times 25$ mas for MJD 51340 and 51341 and $90\times 25$ mas on
MJD 51342.  The vertical lines follow the same tracks as in Figure 1.

    Two estimates of the total flux density for Sco X-1 are shown in
the top two plots of Figure 3.  The flux density, shown in Figure 3a,
was determined from the VLA observations with determination of the flux
density made every 10 minutes.  With a resolution of $3''$ at 5.0 GHz
and $10''$ at 1.7 GHz, these measurements contain the entire emission
from Sco X-1.  Figure 3b shows the total flux density obtained from
the VLBA images.  The good agreement between the VLA and the VLBA
measurements indicates that no significant large-scale structure
associated with Sco X-1 was resolved out by the VLBA observations.
Figures 3c, 3d and 3e show the flux density for each of the three
components.  In Figure 4, the spectral index $\alpha$ of each of the
three components during the dual-frequency VLBA observations is shown.

\subsubsection {A Description of the Changes in Sco X-1}

    The 1999 June observations form a basis for the interpretation of
the radio emission from Sco X-1.  Thus, we will describe the evolution
seen in the images in some detail.

    The basic morphology of Sco X-1 is simple.  The source is usually
composed of three relatively compact components: a radio core that is
nearly coincident with the binary system, a compact NE component
moving away from the core; a weak SW component moving away on the
opposite side of the core.  The three components lie along an axis of
position angle $54^\circ$ to an accuracy of a few degrees.

Changes associated with the emission from Sco X-1 are as
follows:

\begin{itemize}

\item {\bf MJD 51340.1 to 51340.4.} The NE component reached a
maximum intensity, with $\alpha=-0.5$, at MJD 51340.2.  Subsequently,
it faded, became more extended with a steepening spectral index, and
fell below the VLBA detection level at MJD 51340.4.  It did not
reappear over the next ten hours in the images from the APT and EVN.
During this period, the radio core flux density steadily increased with
$\alpha>0.0$.

\item {\bf MJD 51340.4 to 51341.3.} The radio core decreased in flux
density from 20 mJy to 8 mJy in a few hours (the rms error for a flux
density measurement is about 0.5 mJy.)  No emission to the north-east
was present.  At MJD 51340.7 the core flux density began to increase,
and by MJD 51341.0, the flux density had reached 20 mJy.  The images
clearly show that this rise in flux density was associated with a
component emerging from the core to the northeast.  Between MJD
51341.0 to 51341.3, this new NE lobe moved outward and decreased in
flux density.  Emission between the lobe and core was also present.
The core flux density generally decreased in flux density during this
period, but a small flare occurred at MJD 51341.3.

\item {\bf MJD 51341.3 to 51341.8.} After moving about 20 mas from the
core, the NE component brightened, slowly at first, and then abruptly
flared from 2.8 mJy to 20 mJy in less than one hour.  The spectrum
quickly flattened to $\alpha=-0.5$, then NE flux density slowly
decayed to 1.5 mJy over the next 15 hours and its spectral index
dropping to $\alpha\approx -1.2$.  The radio core, during this period,
remained relatively constant at 4 mJy, except for a flare to 8 mJy at
MJD 51342.3

\item {\bf MJD 51341.8 to 51342.2.} The radio core brightened from 4
mJy to 20 mJy, reaching the peak at MJD 51342.0.  The radio core then
decreased in flux density and by MJD 51342.1 extended emission towards
the northeast was clearly present.  However, unlike the core flare and
subsequent expulsion of emission to the NE about 1.3 days earlier,
this extended emission decreased below the detection level by MJD
51342.2 and with no apparent motion.  The SW component, seen better in
the lower resolution 1.7 GHz images, is visible during this period.
The NE component decreased slowly in flux density.

\item {\bf MJD 51342.2 to 51342.4.} For the last four hours of the
observation, the NE component brightened from 1.5 to 4 mJy and the
spectral index flattened again to $-0.5$.  This NE component flare is
less spectacular than the one about 0.8 days earlier, but has similar
properties.  The core remained steady at about 4 mJy with no
resolvable extended structure.

\end{itemize}

\subsection {1998 February 27-28; MJD 50871-2}

   The only previous Sco X-1 observation on consecutive days occurred
on 1998 February 27-28, from two six-hour VLBA observations separated
by 18 hours.  The upper part of Figure 5 shows three sets of
snapshots, each 52 minutes apart on MJD 50871.  The lower two panes
show the structure on the following day when the source was weaker.
Figure 6 shows the flux density of the three components on both days.
Although there was some indication of extended structure between the
core and the NE component, the high-resolution 5.0 GHz images show
that most of the emission was contained in compact components.

   On February 27, the core spectrum was inverted with flux density
about 10 mJy at 5.0 GHz.  The NE component had a flux density $>15$
mJy and was moving from the core in the same position angle as that in
the 1999 June observations.  Near the end of the observation period
the flux density decreased, its spectral index became steeper and the
angular size increased.  This behavior will be discussed in more
detail in $\S$5.7 and with Figure 14.

   Sco X-1 was much fainter on the following day and individual
snapshots were too noisy to show rapid changes.  At the bottom of
Figure 5, the 1.7 GHz and 5.0 GHz images, both with $10\times 5$ mas
resolution, were made from the entire 5-hour observation.  The core
was unresolved with a peak flux density of 3.0 mJy, and decreased
during the day.  The SW component, clearly detected at about 2 mJy,
was also relatively compact.  The radio emission northeast of the core
was complex but confined along the major axis of the radio structure,
between 30 mas to 60 mas from the core at 1.7 GHz.  In the 5.0 GHz
image, only two features in this northeast emission region were
detected above the noise level.  A linear extrapolation of the NE
component motion on MJD 50871 intercepts the most distant component of
emission in the north-east direction on MJD 50872.  Motion of the SW
component was detected on both two days, and is discussed in the
following section.

\subsection {The Component Velocities and Orientations}

   The 1998 February and 1999 June images show the major dynamical
properties of Sco X-1.  A compact component was always present near
the binary system.  It varied significantly in flux density and
occasionally faint emission extended several mas towards the
north-east.  A compact component was usually detected northeast
of the radio core and its motion was clearly directed away from the
radio core.  Its flux density also varied significantly over hour
time scales.  A compact component southwest of the core was detected
about 50\% of the time and was on average about 10 times
weaker than the NE component.

   The relative simplicity of these images of Sco X-1 is contrasted by
the more complicated emission from other well-studied Galactic-jet
radio sources \citep{mir99}.  Although their structures tend to lie on
a well-defined axis, they are often composed of many components on
both sides of the radio core.  For Sco X-1, we find that {\it only one
compact component was detected on the NE and SW sides of the radio
core} (See Figure 11 for a possible exception).

   The changing separation of the NE and SW components from the core
for the 1999 June (top) and 1998 February (bottom) observations is
shown in Figure 7.  For the 1999 June observations, the velocity
in the plane of the sky for the NE component \#1, detected for 6
hours on MJD 51340, was $v=0.73\pm 0.07$ mas hr$^{-1}$ =
0.28c\footnote[1] {For a measured distance to Sco X-1 of 2.8 kpc, one
mas = 2.8 AU = $4.19\times 10^{13}$ cm.  A speed of 1 mas hr$^{-1}$ =
0.388c.}.  The NE component \#2 was not resolved from the core until
about MJD 51341.0.  Then, from MJD 51341.0 to 51341.35, the NE
component speed was $v=1.74\pm 0.16$ mas hr$^{-1}$ =0.68c.  After MJD
51341.4, when the NE component flared, the component speed decreased
to $v=1.25\pm 0.05$ mas hr$^{-1}$ = 0.49c.  Although the NE component
varied considerably in flux density during the last 24 hours of the
observations, its speed remained nearly constant.

   For the 1998 February observations, the velocity of the NE
component on MJD 50871 was $v=1.11\pm 0.06$ mas hr$^{-1}$ = 0.43c over
the five hours of observations, with no significant departure from
linearity except for the last frame.  The single point for the NE
component on MJD 50872 was taken as the position of the most distance
part of the NE component seen in Figure 5.

    Motion of the SW components was more difficult to detect.  On MJD
50872 the component was both strong and long lasting and moving away
from the core with $v=0.6\pm 0.2$ mas hr$^{-1}$.  On MJD 51342 the
velocity was $v=0.5\pm 0.3$ mas hr$^{-1}$.  In Figure 7, the
dashed-lines passing through the positions of the SW components are
not linear fits to its position, but to {\bf 50\% of the fitted
angular separation to the NE component} (Figure 12 shows a plot of
this ratio from all Sco X-1 observations).  Thus, we surmise that the
SW component was also moving away from the core, but at a projected
velocity of about 50\% of that of the NE component.  These motions
are described in more detail in $\S$4.1.

    The orientation of the NE and SW components with respect to the
core for the 1999 June and 1998 February observations are shown in
Figures 8a and 8b.  The radio structure of Sco X-1 remained near
position angle $54^\circ$ during these these two observations, and the
component orientations from the previous observations of Sco X-1 also
lie within a few degrees of $54^\circ$.  Hence, any long-term
variation in the orientation of the radio emission from Sco X-1 is
less than about three degrees.  It is possible to fit the position
angle distribution in 1999 June observations with a mean value of
$54^\circ$ plus a sinusoidal term of $\approx 3^\circ$ with a period
of between two to four days.  Modeling of the binary system X-ray
QPO's suggest that the neutron star/accretion disk interaction could
produce a $5^\circ$ precession over several days \citep{tit00}.

\subsection {The 1995-1996 VLBA Observations at 5.0 GHz}

    The first four VLBA observations of Sco X-1 were made on 1995
August 19 (MJD 49948), 1996 March 16 (MJD 50158), 1996 September 16
(MJD 50340) and 1996 August 3 (MJD 50663), and contour plots of the
radio emission are shown in Figure 9.  These observations had lower
sensitivity than later observations, were only at 5 GHz, and were
about four hours in duration.  The radio emission from Sco X-1 during
these observations was relatively weak so that images made from
one-hour slices of the data were too noisy in order to determine
variability and motion.  Nevertheless, all four observations showed
similar properties in the emission of Sco X-1.
\begin{itemize}

\item {\bf MJD 49948.} The radio core had a peak flux density of 0.32
mJy.  There was a trace of both the NE component and the SW component
19 mas and 12.2 mas from the core respectively, each with a peak flux
density of $<0.20$ mJy.

\item {\bf MJD 50158.} The radio core had a peak flux density of 0.50
mJy.  There was a slight indication of the NE component 14 mas from
the core, with a peak flux density $<0.25$ mJy.

\item {\bf MJD 50340.} The source was stronger during this observation
and all three components were present well above the noise.  The SW
component had an integrated flux density of 2.5 mJy, located 5 mas
from the core.  The core flux density increased from 0.30 mJy to 0.60
mJy during the four-hour observations.  The NE component was about 8
mas from the core with a peak flux density of about 0.6 mJy.  The
distance ratio of the NE and SW components from the core was about
2:1.

\item {\bf MJD 50663.} The core had a peak flux density of 2.0
mJy and the NE component was detected, about 26 mas away, with a peak flux
density of 0.35 mJy.  Any SW component was $<0.15$ mJy.
\end{itemize}

    The relative weakness of Sco X-1 in 1995-1996, compared with later
observations, is consistent with the general characteristics of the
long-term flux density behavior of Sco X-1, as shown from the Green
Bank Interferometer monitoring between MJD 50600 and 51700 (see
\url{www.gb.nrao.edu/fgdocs/gbi/pubgbi/ScoX-1}).  At the
monitoring frequency of 8 GHz, the quiescent level of Sco X-1
remained at 5 mJy for a period of 100 to 150 days, but there were
periods of enhanced emission at 10 to 15 mJy which persisted for 50 to
100 days.  About 5\% of the time there were large outbursts ($>50$
mJy), which lasted over a few days.  These outburst occurred randomly
and were not associated with the periods of enhanced emission from Sco
X-1.

\subsection {1998 August 29-30, MJD 51054-5}

    Figure 10 displays the structure of Sco X-1 for a five-hour
observation on August 29-30, 1998.  Because of the weakness of the
source, each image contains data from the entire five-hour
observation.  All three components were present, although the core was
barely detectable at either frequency.  The NE component was
clearly extended and contained several emission peaks along the source
major axis.  The SW component was compact and relatively strong.  The
appearance of Sco X-1 on this day was similar to that on MJD 50872,
shown at the bottom of Figure 5.  However, the linear size of the
source on MJD 51054 was a factor of three smaller.  Again, the ratio
of the separation from the core of the NE and SW components was about
2:1.

\subsection {1997 August 22, MJD 50682}

    During this five hour VLBA observation, Sco X-1 was extremely
bright and nearly all of the emission was contained within 5 mas of
the core, as shown in Figure 11.  The high-resolution image showed
emission moving outward from the binary system, particularly on the
north-east side of the core.  The relatively bright spot of emission
at the extreme edge of the extended emission was advancing at a
velocity of $1.4\pm 0.3$ mas h$^{-1}$.  Some emission was also
detected SW of the core.  This morphology and evolution was similar to
that seen on MJD 51341.0 when the second NE component was ejected from
the core.  The spectral index of the emission within 5 mas of the core
rose from $\alpha=0$ to $\alpha=0.3$ during the five hour observation.

    A faint component, about 20 mas NE of the core, was also detected.
This is the only example of two radio components on one side of the
core.  However, the distant component is extremely weak with no
measured motion.  The radio structure on this day will be discussed in
more detail elsewhere \citep{gel02}.

\section {Orientation of the Binary system}

\subsection{Doppler-Beaming of NE and SW Components}

    The kinematic results discussed in the section, have already been
reported in Paper I.  We present more detail in this section.

    The images from the 1998 February and 1999 June observations
showed that there was often a compact NE component moving away from
the core.  Although this component varied significantly in flux
density, its velocity remained relatively constant over periods of
many hours.  When the NE component was well-separated from the core,
its measured speeds {\it in the plane of the sky} were 0.28c, 0.43c, 0.49c
and 0.68c.  On MJD 51341 and MJD 50682 a  NE component emerged from
the core.  Because of a lack of sufficient resolution, the emergent
velocity was not well-defined, although it was about 0.5c.

    Motion of the SW component away from the core was observed during
the 1998 February and 1999 June observations, with $v=0.6\pm 0.2$ mas
hr$^{-1}$, and $v=0.5\pm 0.3$ mas hr$^{-1}$, respectively.  Further
detection of the motion was not possible because of the weakness and
intermittent detection of this component.  A comparison of the
NE-to-core and the SW-to-core component distance ratio of all the
observations is shown in Figure 12.  There is no systematic variation
with distance and the average ratio is $0.51\pm 0.02$.  Since the NE
component was moving away from the core, we infer that the SW
component was also moving away from the core in the opposite direction
with 51\% of the velocity of the NE component.

     Another comparison of the radiative properties of the NE and SW
components is given in Figure 13, which displays the spectral index
versus flux density for the two components in Sco X-1 from all of the
observations whenever it was determined.  For comparison, the spectral
properties for the core are shown \citep{gel02}.  When the flux
density of the NE component was stronger than 4 mJy its spectral index
was $\alpha\approx -0.6$ but decreases to $<-0.9$ at fainter flux
densities.  The SW component was detected less than 50\% of the time
and often only at 1.7 GHz.  It greatest flux density was 5 mJy
at 1.7 GHz and the spectral index was about $-0.6$; hence, the flux
density at 5.0 GHz is about half of that at 1.7 GHz.  We conclude that
on average the apparent brightness of the SW component is about ten
times fainter than that of the NE component.  Furthermore, the
spectral index of the SW component at its brightest was about equal to
that of the NE component at its brightest.  We will discuss the flux
density ratio between the two lobes in more detail in $\S$5.9

     The relative positions and velocities of the NE and SW
components, their relative flux density (in a statistical sense since
both components vary considerably), and their spectral properties
suggest strongly that the NE and SW components have similar
properties, with the observed emission differences resulting from
relativistic aberration and Doppler beaming.  The observed apparent
motion in the plane of the sky of a component moving with a velocity v
(usually given in term of c as $\beta=$v/c), at an angle of $\theta$
to the line of sight is \citep{bla77}:

\begin{equation}
\beta_{NE} =  \frac {\beta~\hbox{sin}(\theta)} {1-\beta~\hbox{cos}(\theta)}
\end{equation}
\begin{equation}
\beta_{SW} =  \frac {\beta~\hbox{sin}(\theta)} {1+\beta~\hbox{cos}(\theta)}
\end{equation}
where $\beta_{NE}$ is the observed velocity in the plane of the sky of
the component approaching the observer and $\beta_{SW}$ is the
velocity of the receding components.

  We have measured a range of speeds for the NE component.  However,
the nearly constant position angle of the axis of Sco X-1 in the plane
of the sky (Figure 8) strongly suggests that the direction of the
space velocity of the lobes with respect to the line of sight is also
constant.  To determine an average speed of the two Sco X-1
components, we considered the time ranges MJD 51341.0 to 51342.4 and
MJD 50871.3 to 50872.4.  These periods are instances when the NE
component speed was determined over a long period of time and the SW
component was relatively strong.  These two periods yield a measured
proper motion of $v=1.25\pm 0.05$ and $v=1.11\pm 0.06$ mas hr$^{-1}$.
We will thus assume that the average speed of the NE component of Sco
X-1 of $1.18\pm 0.08$ mas hr$^{-1}$.  At a distance of $2.8\pm 0.3$
kpc for Sco X-1, the average speed of the NE component in the plane of
the sky is then $\beta_{NE}=0.46\pm 0.08$, with the distance
uncertainty incorporated in the error.  Next, we adopt the ratio of
the SW speed to the NE speed as $0.51\pm 0.02$.  This ratio is a
function of the speed, but should be the appropriate value that we
have adopted.  Equations (1) and (2) become:

\begin {equation}
\frac{\beta_{SW}}{\beta_{NE}} = 0.51\pm 0.02 = \frac{1-\beta~\hbox{cos}(\theta)}{1+\beta~\hbox{cos}(\theta)},
\end {equation}
\begin {equation}
\beta_{NE} = 0.46\pm 0.08 = \frac{\beta~\hbox{sin}(\theta)}
{1-\beta~\hbox{cos}(\theta)},
\end{equation}
from which we obtain
\begin {equation}
\beta = 0.45\pm 0.03;~~~\theta = 44^\circ \pm 6^\circ.
\end {equation}
The quoted errors are the one-sigma uncertainties and include the
measurement errors.

    The determination of the Doppler boosting associated with the flux
densities of the NE and SW components requires a specific model of
their internal structure, particle motion and magnetic field geometry
as discussed in $\S$5.5.  If we assume that the moving components
are transparent, with all particles moving at the bulk velocity and
radiating isotropically, then the flux density ratio, $R$ is given by
\citep{bla79, caw91}:
\begin{equation}
 R= \left[\frac{\beta_{SW}} {\beta_{NE}}\right]^{k-\alpha}
\end{equation}
where k is a geometric factor which equals 2 for an optically thin
source, and $\alpha$ is the spectral index of the source emission.
From the measured $\alpha=-0.6$ we obtain a predicted flux density
ratio of R=0.17, which is somewhat larger than that observed (see
$\S$5.9 for more details).  Thus, the space motion derived solely
from the relative motions of the two components is in reasonable
agreement, considering many uncertainties about the radiation
properties and variability of the lobes.

\subsection {Sco X-1 Binary Masses}

    Assuming that the radio beam lies along the rotation axis of the
accretion disk which in turn lies in or close to the orbital plane of
the binary system, we can better determine some of the properties of
the binary.  The relative constancy of the radio axis over five years
are supports the above association.

     The mass function for the Sco X-1 binary has been derived from
the measured radial velocity of the HeII $\lambda$-4686 line,
presumably from the accretion disk, and is given by $F(M)$ =
$(M_2~\hbox{sin}~i)^3 / (M_1+M_2)^2$ = $0.016\pm 0.004$ (our estimated
error), where $M_1$ and $M_2$ are the masses of the neutron and
secondary star, respectively, and $i$ is the inclination of the orbit
\citep{cow75}.  The binary is composed of a degenerate star, almost
certainly a neutron star \citep{has89}.  Measurements of many neutron
star systems produced estimated masses between 1.2 to 1.6 $M_\odot$
\citep{lew95}.  The intrinsic X-ray luminosity of Sco X-1, for a
distance of 2.8 kpc, also suggests a neutron star mass of about
$1.4~M_\odot$.  The calculated secondary mass $M_2$,
from the best value of the mass function, inclination and neutron star
mass, is ($0.63\pm 0.26)~M_\odot$.  Such a low companion mass is a
problem for most models of the energetics in Sco X-1.  First, the star
may not fill its Roche lobe resulting in an accretion rate that is too
low; perhaps the companion is an evolved star.  Secondly, infra-red
spectroscopic observations of the binary system suggest that the
secondary star is earlier than G5 \citep{ban99}.  However, the X-rays
from Sco X-1 in all likelihood heat up the atmosphere of the
secondary, puffing out its atmosphere until it overflows the
Roche lobe reulsting in mass accretion onto the compact star.

    The secondary star mass estimates can be increased in the several
ways.  If the degenerate star mass is $>3.5~M_\odot$, then the
secondary star would have a mass $>1$ M$_\odot$.  But, then what is
the nature of the degenerate star?  A mass function $>0.025$ would
also increase the secondary star mass associated with a $1.4~M_\odot$
neutron star.  This implies that the optical lines are coming from a
much larger region than the accretion disk near the neutron star.
Thus, the masses of the neutron star and the secondary member of the
Sco X-1 binary are still uncertain even with knowledge of the orbit
inclination.

\section {The Nature of the NE Component}

\subsection {Angular and Linear Size}

    The angular size of the NE component could be measured with an
accuracy of 0.5 mas when it had a flux density $>5$ mJy.  Figure 14
shows a plot of the angular size and motion of the NE and core
components for the seven snapshots on MJD 50871 (see Figure 5 also)
when both the core and NE components were relatively strong.  The
angular sizes were determined by fitting the visibility data directly
to an elliptical Gaussian model and by fitting the component
brightness on the image and deconvolving the effects of the beam
shape.  Both methods gave consistent results.

    The comparison of the positions and sizes of the core and the NE
components provides a good indication of the robustness of these
measurements.  The core was about 10 mJy, relatively stationary and
extended in the direction along the source major axis, as was commonly
observed at other epochs.  Alternatively, the NE component was
extended orthogonal to the direction of motion and was clearly moving
away from the core.  Since the effects of noise and poor phase
calibration will blur all of the emission in the same way, the
different size and orientation of the core and NE components strongly
suggests that these measured angular sizes and orientations are valid.

    For snapshots 1 through 3 in Figure 14, the measured average
angular size and orientation of the NE component itself was ($1.5\pm
0.15) \times (0.9\pm 0.2$) mas in position angle (pa) $140^\circ\pm
15^\circ$.  The angular size in snapshot 4 is marginally bigger and
the size for snapshot 5 is $2.4\times 1.3$ mas in pa $120^\circ$.  In
the 50-minute separation between snapshot 5 and 6, the component
increased in area by a factor of 4.5, to $5.0\times 2.9$ mas in pa
$150^\circ$, an expansion at a velocity close to c.  In snapshot 7 the
size of the component did not increase substantially, but the bulk
motion of the emission may have decreased.  The angular size of the NE
component was also accurately determined on MJD 51341.4 and MJD
51342.3.  The angular sizes were ($1.6\pm 0.1) \times (0.7\pm 0.3$)
mas in pa $135^\circ\pm 10$ and ($2.4\pm 0.3) \times (0.8\pm 0.4)$ ma
in pa $155^\circ\pm 30^\circ$, respectively, in good agreement with
the minimum size of the NE component on MJD 50871.

    The NE component is clearly extended in the direction
perpendicular to its motion.  In addition, the measured size in this
direction of motion is slightly increased by the component motion
during each snapshot over a 33-minute span.  During this period, the
component moves 0.6 mas; hence, the measured size of 0.9 mas
corresponds to a true size of about 0.7 mas after removing the
blurring from the motion.  Thus, we derive a minimum component size
(full-width at half power) of $1.5\times 0.7$ mas, corresponding to
$6.3\times 10^{8}$ km by $2.9\times 10^{8}$ km with the major axis
perpendicular to the orientation of the source.  More detailed
measurement of the NE component structure, other than its angular
size, cannot be obtained with the present resolution and SNR.

  The flux densities and spectral indices for the NE components are
also listed in the Figure 14.  The relationship between the angular
size, flux density and spectrum of the NE component are discussed in
$\S$5.7.

\subsection {Component Energetics}

     The calculation of the energetics of the NE component used the
following parameters: a source distance of 2.8 kpc ($8.6\times
10^{16}$ km), a flux density at 1.7 GHz of 30 mJy, $\alpha=-0.6$, with
the radio cutoff frequencies of $10^7$ to $10^{11}$ Hz.  The volume of
the component (ellipsoid of diameters $1.5\times 1.5\times 0.7$ mas
with the narrow dimension in the direction of the component advance)
is $6.1\times 10^{40}$ cm$^3$.  The integrated flux density is
$6.5\times 10^{-21}$ erg s$^{-1}$cm$^{-2}$, the total observed radio
luminosity is $6.1\times 10^{30}$ erg s$^{-1}$.  We assume a factor of
3 to correct for the Doppler boosting of the NE component (and Doppler
attenuation of 3 for the SW component) as implied by the average flux
density ratio of the two component .  The radio luminosity in the
frame of the source is $2.0\times 10^{30}$ erg s$^{-1}$, corresponding
to a brightness temperature of about $1\times 10^9$ K.  The total
energy in relativistic particles $E_{e}$, and total energy in the
magnetic field $E_{H}$, assuming the above radiating volume, are
\begin{equation}
E_{e} (\hbox{ergs}) = 4.0\times 10^{37}~H^{-1.5}
\end{equation}
\begin{equation}
E_{H}(\hbox{ergs}) = 2.4\times 10^{39}~H^2,
\end{equation}
where $H$ is the magnetic field in Gauss.

    The radiative lifetime of the emitting particles is about one year
if magnetic field energy and kinetic energies are about equal.  For
example, if we assume that the protons contain 100 times the energy as
the electrons, the equipartition magnetic field is about 1.0 G with a
minimum total energy of $6\times 10^{39}$ erg in the component.  If
the electron energies dominate, then the minimum total energy is about
$5\times 10^{38}$ erg and has a field strength of about 0.3 G.  Such a
long radiative lifetime is not possible for the NE component since it
is variable on an hour time-scale and a size of $5\times 10^8$ km
(30-min light travel time).  This problem will be discussed in
$\S$5.6.

\subsection {What is the NE component?}

     Emission confined to relatively discrete components, and which
show relativistic motion have been observed in many Galactic X-ray
binary systems \citep{mir99}.  Three explanations are generally given
for the energetics of these components: (1) Ejected clouds, where the
components are clouds of radiating particles, expelled from the binary
in a preferred direction, but containing their own energy source; (2)
Interaction within beams; where the energy flow within twin-beams is
collimated by an accretion disk of the massive object and interacts
with embedded matter or entrained material to produce radiating
electrons; (3) a working surface where the energy flow in the
efficient twin-beams is relatively invisible until the flow impinges
on external material and forms a bright, small region of highly
relativistic particles.  The beam also penetrates through the external
medium at relativistic velocities.  We suggest that the moving
components on opposite sides of Sco X-1 are regions of intense radio
emission, generally denoted as {\it lobes}, formed from the
interaction of a twin-beam with the ambient medium in a confined
region called a {\it working surface}.

     There are two properties of the components in Sco X-1 that that
are best explained by this lobe model.  First, the rapid variability
and lack of systematic decay of these components implies that the
radiating electrons have a lifetime of less than one hour and are
constantly resupplied with energy.  In fact, the NE component
variability is similar to that of the core.  In contrast, the emission
from the relativistic moving components associated with the other
well-studied Galactic-jet radio sources are not as variable and tend
to decrease fairly uniformly as they move away from the radio core
\citep{fen99, hje95b}.  Secondly, the observations of Sco X-1 suggest
that only one pair of components (NE and SW) occur at any given time
(see Figure 11 for a possible exception).  The core does show activity
and extended emission towards the northeast direction, but more than
one compact component is not observed on either side of the core.  For
the other Galactic-jet sources, several components are usually visible
on one side of the core and they are often associated with X-ray
and radio flaring events in the core region.

     However, the later stages of the lobe evolution in Sco X-1 did
show a more complicated structure.  For example, the radio emission
from Sco X-1 on MJD 50872 and MJD 51054 had an extended NE component
containing several bright regions.  On MJD 50872 this extended NE
component is clearly associated with the moving compact component
observed on the previous day and consists of an extended lobe towards
the core with a few small brighter regions.  Perhaps this extended
emission is back flow from the lobe region seen in many extragalactic
objects \citep{bla92}.

\subsection{The Advance Velocity of the Lobe}

     The speed of advance for three different lobe-pairs were measured
over the range 0.32$<\beta<$0.57.  The NE lobe on MJD 51340 moved at
0.32c just before evaporating.  The lobe on MJD 50871 moved at 0.43c.
The lobe on MJD 51341-2 moved initially at 0.57c, but abruptly
decreased to 0.46c just after a lobe flare.  The advance speed during
each epoch was remarkably constant over periods of many hours, up to
one day.

     A prediction of the advance speed of the lobe requires a detailed
model of the interaction of the beam with the ambient medium, A more
general approach \cite{bla79} has derived the velocity of the advance
of a lobe under the assumption that the relativistic shock produces an
ultra-relativistic equation of state on {\it both} sides of the shock.
This region is generally called the {\it working surface} and this
region produces a radio hot-spot which may be a small, but intense,
part of the lobe.  If $p_{1}$ and $p_{2}$ are the pressures and
$v_{1}$ and $v_{2}$ are the velocities in the unshocked (beam) and
shocked (working-surface) regions, then
\begin{equation}
\frac{v_1}{c}= \sqrt{\frac{p_1+3p_2}{3(3p_1+p_2)}}~~~~\frac{v_2}
{c}= \sqrt{\frac{3p_1+p_2}{3(p_1+3p_2)}}.
\end {equation}
When the beam pressure is much greater than the shock pressure, the
advance speed is 0.33c.  For more dominant shock pressure, the advance
speed eventually reaches c.  This model gives results which are
consistent with the range of observed speeds using reasonable pressure
ratios across the hot-spot.  However, the small variation of the
observed speed with time is still puzzling.

\subsection {Lobe Emission Geometry}

     The ultra-relativistic fluid produced at the working surface
fills the larger volume of the radiating lobe.  Figure 15 shows a
schematic of the general appearance of the lobe based on some very
simple assumptions.  There are two velocities associated with the
lobe: the average bulk motion velocity of $\approx 0.45$c, and a
diffusion velocity of 0.57c associated with this relativistic gas.  We
assume further that the radiative lifetime of the emitting particles
is about 30-minutes as required by the variability time-scale.  The
\lq wagon-wheel\rq~diameter (in the frame of the lobe) is then
$6.2\times 10^{8}$ km in diameter, and the thickness is the distance
traveled by the lobe in 30 minutes, $2.4\times 10^{8}$ km.  The lobe,
however, viewed at an angle of $44^\circ$ to the line of sight, would
make the component appear somewhat more circular than the above model.

    A deeper analysis of the physical processes in the lobes requires
detailed hydrodynamic jet modeling and the radiative transfer of the
electrons.  The modeling of \cite{mio97}, for example, shows striking
similarities between the structure of Sco X-1 with some of their
simulated models at a viewing angle of $45^\circ$ to the line of sight
(see their Figures 1d, 5c, 16a).  The predicted peak brightness ratio
between receding and advancing lobes is R=0.05.  However, this model
assumes an isotropic magnetic field distribution.  In any case, there
seems to be a small family of models whose parameters can be chosen to
match the morphology as well as the physical properties and
characteristics of the lobes.

\subsection{Lobe Emission Lifetime}

     The lifetime of the radiating plasma in the lobe cannot be much
longer than 30 minutes, the variability time-scale.  The two most
likely sources of energy depletion are from synchrotron losses or from
adiabatic expansion of the plasma.  First, if equipartition holds
between the magnetic energy and particle energies as discussed
in $\S$5.2, then
the time scale for synchrotron losses is about one year for the
equipartition field strength of about 0.5 G.  The radiative lifetime
decreases with as H$^{-1.5}$; hence a one hour decay time requires a
magnetic field strength of H=300 G.  The derived lobe energies are
then
\begin {equation}
E_H\hbox{(ergs)} = 2\times 10^{44} \hbox{ for H= }300~G
\end{equation}
\begin {equation}
E_{e}\hbox{(ergs)} = 7\times 10^{33} \hbox{ for H= }300~G.
\end{equation}
The magnetic energy is then a factor of $10^8$ larger than the
equipartition value, even assuming a density of protons about 100
times that of the electrons.  Magnetic field amplification just in
advance of the shock has been suggested \citep{dey80,hug81}, although
not to this extreme.  With a lobe brightness temperature $<10^9$K, the
rate from inverse-Compton losses is much less than for synchrotron
losses even with this large magnetic field.  Such a large magnetic
field requires a lobe energy density which is about $10^5$ times
larger than the minimum energy value.  The outer part of the radio
lobe would also have a steeper radio spectrum than near the working
surface and the 1.7 GHz angular size would be larger than that at 5.0
GHz.  Unfortunately, the present observations just barely resolve the
lobe at 5.0 GHz and such a spectral gradient is not observable.

     The radiating electrons generated within the working surface can
also rapidly lose energy if they expand adiabatically to fill the lobe
region.  Assuming that the radio emission is optically thin (the radio
spectral index $\alpha\approx -0.6$ is consistent with this) and
neglecting the motion of the lobe for the moment, the steady-state
emission profile of an expanding spherical cloud which contains a
steady source of accelerated electrons within a radius $r_0$, the
working surface region, is
\citep{bal93, hje95a}
\begin{equation}
         S(r) = S_0(r/r_0)^{4\alpha-1}
\end{equation}
where $S_0$ is related to the inserted energy per unit time in the
working surface volume, $r$ is the distance from the center of the
working surface.  For $\alpha=-0.6$, the intensity at $r/r_0=1.22$, is
already less than half of the intensity at the working surface.  Hence
the adiabatically expanding electrons need not diffuse far from the
working surface in order to lose most of their energy.  From the
measured lobe size of about 1 mas and a diffusion speed of 0.57c, the
electrons will lose most of their energy in about 40 minutes.  The
bulk motion of the lobe is somewhat less than the diffusion speed and
does not substantially change the above calculation.  Adiabatic
expansion does not modify the spectral index of the emission across
the lobe, in contrast to the spectral steeping with synchrotron
ageing.  In the case of pure adiabatic expansion, the working surface
is an appreciable part of the observed radio lobe.

    A combination of both loss mechanisms may be operating in the
lobes of Sco X-1.  For example, the volume close to the working
surface may have a large magnetic field which completely dominates the
plasma dynamics.  Further away from the working surface, adiabatic
expansion may be the dominent loss mechanism.  This would reduce the
total energy content needed to supply the lobe emission compared with
that needed if only synchrotron losses were operating.

\subsection {Lobe Expansion}

    A slow expansion of the lobe angular size over an hour or two
time-scale was observed several times during the observations of Sco
X-1.  The best example of the expansion of a lobe was on MJD 50871
after frame 3 in Figure 14.  Between frames 5 and 6 the angular size
increased by a linear factor of 2.2 in 50 minutes.  As discussed in
the above section, for pure adiabatic expansion with a uniform input
energy source, the flux density in frame 6 should have decreased
$(2.2)^{4\alpha-1}$ \citep{hje88} or about 4\% of that in frame 5.
However, the flux density at snapshot 6 was 60\% of that in snapshot
5.

     Without higher resolution observations of the lobe we can only
speculate on this slow lobe expansion phenomenon.  Since adiabatic
expansion alone would produce a much more rapid decrease of flux
density than that observed, we suggest that the increasing size and
decreasing intensity of the lobe reflects the energy flow and size of
the working surface.  The somewhat decreasing spectral index could be
related to the acceleration process in the working surface, or the
effect of synchrotron ageing.  Note that this lobe, although
decreasing in flux density at the end of MJD 50871, is present on MJD
50872 as an extended lobe along the axis of the source.  Hence, this
lobe was reinvigorated sometime after the end of the observations on
MJD 50871.  Much more data are needed in order to explain the general
appearance and evolution of these components.

    A similar expansion and spectral steepening of a lobe occurred
during the first four hours on MJD 51340.  From the images in Figure
2, the component clearly expanded, the flux density decreased and the
spectral index steepened to $-1.5$ until the component fell below the
detection level of the observations.

\subsection {Energy Flow Velocity in the Beam}

    Paper I we showed that the the flux density variations of the radio
components have good temporal correlation when assuming a model of
energy flow from the core to the lobe where: (1) a core flare is
associated with an event near the binary system (2) a burst of energy
associated with this event travels down the twin-beam at a velocity of
$\beta_{j}$, (3) this increased energy flux intercepts the NE and SW
lobes and produces a flare.

    The appropriate time delays between the NE and SW lobes with
respect to the observer/core reference frame can be determined using
the known lobe velocity $\beta$ and orientation $\theta$ to the line
of sight.  This model predicts the delay of the lobe flares
$\tau_{NE}$ and $\tau_{SW}$, with respect to the driving core flare as
viewed by the observer, to be:
\begin {equation}
\tau_{NE} = (t_1-t_0)\beta
\frac{(1-\beta_j\hbox{cos}\theta)}{(\beta_j-\beta)},
\end {equation}
\begin {equation}
\tau_{SW} = (t_1-t_0)\beta
\frac{(1+\beta_j\hbox{cos}\theta)}{(\beta_j-\beta)}.
\end {equation}
Here $t_0$ is the time of ejection of the lobes from the core, and
$t_1$ is the time of a core flare.  With this model, the NE lobe should
flare at time $t_1 + \tau_{NE}$ and the SW lobe at time $t_1 +
\tau_{SW}$ later than the core, as viewed by the observer.  The term
$(t_1-t_0)\beta$ is the true distance of the lobes from the core at
time $t_1$.

    Figure 16a shows the observed flux density versus time for
the three components.  Figure 16b shows the flux density
correlations after correcting for the model delay, assuming
$\beta_j=1$.  Figure 16c shows the same data assuming $\beta_j=0.90$
and the correlations between the core with the NE and SW lobe flux
density flares are significantly worse.  Additional discussions of the
correlations and its relationship to various properties of Sco X-1
were presented in Paper I.

     A lower limit to the flow speed in the beam, as indicated by the
difference in correlation between Figures 16b and 16c, can be
determined more quantitatively.  It seems unlikely that the apparent
correlations in Figure 16b are pure chance.  There are three pairs of
flares (C3-L3, C4-L4, C3-S3) which are clearly correlated; all are
associated with a significantly different delay, using a model with
only one free parameter, derived solely from the positional
information of the components.  The correlation of the flares for
$\beta_j=1.0$ is clearly better than that for $\beta_j=0.9$ (and by
symmetry for $\beta_j=1.1$), but how much better?

    An estimate of the uncertainty can be obtained by determining the
speed necessary to align each of the three flare correlations (C3-N3),
(C4-N4) and (C3-S3).  These results yield the three independent
estimates of $\beta_j$ of $1.10\pm 0.10$, $1.00\pm 0.04$, and $1.02\pm
0.04$ respectively, which gives a weighted average $\beta_j=1.02\pm
0.04$.  Changing the space velocity of the lobes within the estimated
uncertainty affects $\beta_j$ only at the few percent level.  The
relatively large beam speed of 1.10c associated with flare (C3-N3) may
be caused by an additional delay in the flare of the core.  We have
suggested that the core may be at the base of the jet, at a distance
of one or two mas from the binary.  This would cause a delay in the
core flare reaching the observer by $\approx 0.1$ hours, compared the
event in the binary system.

     Thus, we derive a lower limit of $\beta_j>0.95$ based on the
above analysis.  Although the analysis indicates that this limit is
significant at the two-sigma level, the model may be too simplistic.
We believe that the limit has a confidence level of 70\% (one-sigma).

\subsection {Flux Density Variations and Doppler-Boosting}

     Based on the space motion of the lobes, we calculate that Doppler
boosting will produce a flux density ratio of the SW lobe to the NE
lobe of R=0.17 assuming a uniformly filled lobe with an isotropic
magnetic field.  The observed flux density ratio is difficult to
determine because of the variability of the lobes and the time delay
in their reference frames.  However, an estimate can be obtained by
two methods.  First, Figure 13 shows the distribution of the flux
densities of the NE and SW components over all of the observations.
At 5.0 GHz the highest flux densities observed for the NE component
are about 18 mJy.  For the SW component, the largest flux densities
are 4.5 mJy at 1.7 GHz, which corresponds to about 2.2 mJy at 5.0 GHz
assuming $\alpha=-0.6$.  This ratio is R=0.12.  A discussion of the
core and its relationship to the x-ray and optical emission is given
elsewhere \citep{gel02}.  Second, another estimate of the
Doppler-boosting can be made from the flare peak \#3 for the NE lobe
and the SW lobe, shown in Figure 16.  The increased flux densities at
5 GHz in the flare, compared with the baseline derived from a few
hours before and after the flare, are: Core flare = 3.5 mJy, NE lobe =
17.0 mJy, and SW lobe = 1.0 mJy.  The ratio between the SW and NE lobe
flare height is R=0.06.  This ratio assumes that the energy flow in
the twin-beams are the same.  Finally, hydrodynamic modeling of a
source with similar properties as Sco X-1 \citep{mio97} gives suggests
a ratio of R=0.07 for the peaks of the simulated lobes--which is
resolution dependent.  All of these estimates are lower than the
R=0.17 expected purely from the space motion; however, any complexity
in the emission properties of the lobes (eg. internal structure,
non-isotropic magnetic field) tends to decrease this expected ratio
\citep{caw91}

\subsection{Comparison with other Galactic-Jet Sources}

     Why does Sco X-1 show a lobe phenomenon when the observations of
other well studied Galactic-jet sources generally show multiple
components?  The general consensus from the other well-studied
Galactic-jet objects like GRS1915+105 (Dhawan et al.\ 2000, Fender et
al.\ 1999), GRSJ1655 (Tingay et al.\ 1995; Hjellming \& Rupen 1995),
CygX-3 (Geldzahler et al.\ 1984; Marti et al.\ 1992. Schalinski et
al.\ 1993; Mioduszewski et al.\ 2001), SS433 (Margon 1984; Spencer
1984), and Cir X-1 (Fender et al.\ 1998), is the radiating material is
probably associated with discrete clouds or shocks being energized
within a collimated flow of energy.  Most of these objects are
transient X-ray sources, preferentially observed just during or after
a strong flare.  These flares may be associated with the ejection of a
large amount of material that is entrained in a collimated flow of
energy, producing strong radio emission.  However, there are several
periods when the core of Sco X-1 showed changing radio emission
usually within 5 mas towards the northeast.  Two of these events are
associated with the emergence of the new NE component on MJD 51341.0
(Figure 2) and MJD 50682.0 (Figure 11), and have already been
discussed.  After the core flare at MJD 51341.95 (Figure 2), extended
emission to the NE persisted for about four hours.  A compact NE
component more than 15 mas from the core was also present.  Figure 14
shows that the core on MJD 50871 was also slightly extended by about 3
mas toward the north-east direction.  The extended emission near the
core may be associated with material in the beam flow.

   The degenerate object associated with Sco X-1 is probably a neutron
star, although the improved orbital parameters, based on the
orientation of the source discussed in $\S$4.2, do not confidently
constrain the degenerate star mass to less than about 2 M$_\odot$.
For most other Galactic-jet sources, the degenerate star is probably
too massive to be a neutron star.  Sco X-1 is also a persistent X-ray
source, unlike many of the other well-studied Galactic transient X-ray
sources.  The relationship between the X-ray properties with the
nature of the degenerate star and with the properties of the radio
emission is still unknown.

   Finally, the major difference between Sco X-1 and the other sources
may be the viewing angle of the source as seen by the observer.  The
simulated images \citep{mio97}, which showed a radio structure
structure surprisingly similar to that seen in Sco X-1, take on the
appearance of the other Galactic-jet sources when the viewing angle is
decreased.  At small viewing angles, the shocked material moving near
c are greatly Doppler boosted and dominate over the emission from the
more slowly-moving lobes.  Thus, like the extra-galactic sources, the
radio appearance of Galactic-jet sources may also be a strong function
of viewing angle.

\section{Comparison with Extragalactic Sources}

    One of the major reasons for studying Galactic-jet sources is to
understand better the jet phenomenon in extragalactic radio sources.
The appropriate scaling laws come from considerations of the accretion
disks and infalling mass \citep{beg84}.  The radio emission is a
second order product of the accretion process, but should also follow
the scaling.  The luminosity ratio of Sco X-1 to Cygnus A, for
example, is $10^{-7}$.  The mass ratio of the neutron star in Sco X-1
and the massive object in Cygnus A is $10^{-9}$.  We will adopt as a
typical scaling parameter $x$, from a radio galaxy to Sco X-1
$x\approx 10^{-8}$.  The luminosity, the time scale and the size will
scale as $x$, the density as $x^{-1}$, the magnetic field as
$x^{-0.5}$, and velocity and temperature are invariant to first order.
The following are some examples of simple scaling of the properties of
Sco X-1's radio lobes to those in extragalactic radio sources.

    Two of the best studied extragalactic radio sources with strong
lobes are Cygnus A \citep{car91} and Pictor A (Perley et al.\ 1997).
Comparisons of these and other sources with Sco X-1 indicates:

\begin{itemize}

\item {\bf Lobe Variability.} The lobes in Sco X-1 vary on an hour
time scale.  This scales to $1\times 10^4$ yr for radio galaxies.
Variability in extragalactic lobes have not yet been observed over
about a 30-year span.

\item {\bf Lobe Lifetime.} The lifetime for a pair of lobes
in Sco X-1 is about one day.  This scales to $5\times 10^5$ yr for a
typical extragalactic source, and is about the estimated age of the
Cygnus A hot spots, based on the curvature of their radio spectrum
\citep{car91,cox91}.

\item {\bf Lobe Regeneration.} In Sco X-1 there appears to be
replenishment of the radio lobes.  That is, after a pair of lobes
fades away, another pair emerges from the core, probably in less than
one day.  The one well-documented disappearance of the lobes (MJD
51840) occurs after an X-ray flare and the emergence of an
optically-thick core component.  The suggestion that the the beam flow
has been disrupted and later reforms is discussed elsewhere
\citep{gel02}.  Although a short lifetime of $10^5$ yr for hot-spots
in radio galaxies has been suggested from lobe modeling \citep{cox91}
and observations of radio galaxy lobes \citep{bla92}, new hot-spots
are believed to appear in the established radio-lobe region rather
than being reformed near the core.  Hence, unlike Sco X-1, the beam
flow in extragalactic sources may not be completely disrupted.

\item {\bf Lobe Size.} The linear size of the lobe in Sco X-1 of
$2\times 10^{8}$ km scales to 500 pc for radio galaxies, similar to
that observed in radio galaxies \citep{bla92}.  The western lobe in
Pictor A \citep{per97} is well-resolved and is somewhat more elongated
perpendicular to the presumed input energy flow, as found for Sco X-1.

\item {\bf Lobe Distance from Core.} The lobe separation of Sco X-1 is
about 40 mas or $1.6\times 10^{10}$km.  This scales to a linear size
of $1.6\times 10^{18}$km or 50 kpc for the typical size of a radio
galaxy.

\item {\bf Spectral Index.} The spectral indices associated with the
lobes of Sco X-1 are similar to those seen in extragalactic sources,
where the most confined lobes have $\alpha\approx -0.5$ and the less
confined have steeper spectral index.

\item {\bf Opening angle.} The ratio of the lobe size to the distance
from the core is probably related to the angle opening in the beam.
For Sco X-1 this angle is typically about $1^\circ$.  A similar ratio
is observed for extragalactic double sources that have prominent
hot-spots in their radio lobes.

\item {\bf Magnetic Field Strength.} The equipartition magnetic field
in Sco X-1 was $\sim 1$ G.  If adiabatic expansion of the plasma in
the lobes is not occurring, the field needed to decrease the electron
emitting lifetimes from synchrotron losses to one hour was $\sim 300$
G.  These two magnetic fields strengths scale to $1\times 10^{-4}$ G
(which is similar to the equipartition field in Cygnus A) and $3\times
10^{-2}$ G respectively.  The configuration of the magnetic field in
the lobes of Pictor A appears to be circumferential \citep{per97}.  The
degree of polarization in Pictor A is also high, indicating an ordered
field.  No lobe polarization information is available for Sco X-1.
However, if the lobes are dynamically controlled by a large magnetic
field, a radial magnetic field would be more likely with Sco X-1.

\end{itemize}

    A major difference between the radio lobes of Sco X-1 and those
associated with radio galaxies is the apparent simplicity of the Sco
X-1 lobes, usually consisting of two hot-spots and no extended
emission.  In contrast, radio galaxies often show an extended lobe,
with a cocoon of emission, and a hot-spot.  Emission from the beam
between the core and the lobe is also observed in some cases.  Often a
secondary hot-spot is observed in a radio lobe, not far from the more
compact one.  These secondary hot-spots are thought to be caused by
either the splatter of material from the primary hot-spot, or the
remnants of older hot-spots which are still radiating \citep{cox91}.

     The difference in appearance between Sco X-1 and a typical radio
galaxy is {\it not} the result of the high dynamic range images
available for many extragalactic sources.  Even at the modest dynamic
range of the Sco X-1 images (peak flux density to rms noise level of
about 50) nearly all double radio sources have more complicated lobes
than that observed in Sco X-1.  Perhaps the suggested strong magnetic
field in Sco X-1 confines the relativistic fluid to a small region.
In contrast, on MJD 50872 and MJD 51054 the NE lobe of Sco X-1
resembled the morphology of that in FRII-type radio galaxies.  The
lobe was extended along the source axis and the hot-spot was not
dominant.  For the MJD 50872 observations, we know that the lobe was
formed more than 1.5 days earlier, suggesting that the confinement of
relativistic particles within a small region is less likely as Sco X-1
evolves.

    Explanations for the differences seen between the radio lobes in
Sco X-1 and those of extragalactic sources require a better
understanding of the formation of the working-surface and the
relativistic particles and fields and their transport into the
hot-spot.  One significant difference between Sco X-1 and radio
galaxies may be the advance speed of the hot-spot.  For Sco X-1, the
advance velocity is 0.5c. It is believed that the advance velocity in
radio galaxies is only mildly relativistic $\approx 0.1$c.  Another
difference may be that the lobes in extragalactic sources are probably
confined by ram pressure.  In Sco X-1, the lobes may be confined by a
strong magnetic field, or, not confined at all with adiabatic
expansion occurring outside of the working surface region.

\section  {Conclusion}

   These extensive VLBI observations of Sco X-1 are among the most
detailed for a Galactic-jet radio source.  The high-resolution images
over the four-year study show relativistic motion and variability of
all components and a recurrent phenomenon of lobe production.  The
resolution and sensitivity are sufficient to determine accurate
parameters of the discrete components as a function of time.  Since we
have observed Sco X-1 without regard to its X-ray and radio flux
density, we may be sampling the more stationary properties of a
Galactic-jet source rather than those associated with the evolution
during the more explosive periods.

   The relatively non-speculative conclusions from these observations
are the following.

\begin{itemize}

\item Sco X-1 is composed of three components: a radio core
near the compact component, a NE component which moves away
from the core, and a weaker SW component which is detected about half
of the time.

\item Occasionally weak extended emission emanates from the core
in the north-eastern direction, but generally only a NE and SW
component are detected.

\item All components are variable on time scales of about one hour.
The rapid increase and somewhat less rapid decrease in the flux
density for all of the components has a characteristic time scale of
about three hours.  This time-scale of the radio variations in Sco X-1
has already been noted \citep{hje90, bra97}.

\item A pair of moving components persists up to about two days.
Recurrent generation of pairs occurs often enough that a naked core
component is not commonly observed.

\item The comparison of the radio properties between the NE and SW
components strongly suggest that they are intrinsically similar, but
differ from the effects of relativistic aberration and Doppler
beaming.

\item The average speed of the NE and SW components is $0.45c\pm
0.03c$, radially away from the core at an angle $44^\circ\pm 7^\circ$
to the line of sight.

\item The component speeds remain unchanged for many hours although
the speeds for different pairs of components at different times range
between 0.31c and 0.57c.

\item The components of Sco X-1 lie on a axis which has not
varied by more than about $4^\circ$ over five years.  A
variation of axis direction of $\approx 3^\circ$, with a period of a few
days, may be present.

\item The minimum angular size of the NE (and presumably the SW)
component is $1.5\times 0.7$ mas ($6\times 3 10^13$ cm) with an
orientation {\it perpendicular} to the motion of the lobe.  This
corresponds to a volume of $6.1\times 10^{40}$ cm$^3$.  The
equipartition magnetic field is between 0.3 to 1.0 G depending on the
ratio of electrons to protons.

\item The lobes of Sco X-1 are dominated by a compact hot-spot.  The
lobes of most extragalactic sources often contain one or more hot-spots
and much more extended structure.

\end {itemize}

     More speculative conclusions are:

\begin{itemize}

\item The NE and SW components are lobes generated from a working
surface where the energy flow within a twin-beam from the binary
impinges on the ISM.  This conclusion is based on the variability of
the components, the lack of a systematic decrease in the lobe flux
density with time, and the simplicity of the components.

\item If the above suggestion is correct then the temporal
correlation of the core flare times with those of the NE lobe and SW
lobe suggest a beam velocity $>0.95$c.

\item In Sco X-1 the lobe emission dominates.  In most other
Galactic-jet sources, emission from material or shocks within the jets
dominate.  This may be an intrinsic difference between the sources,
perhaps related to either the nature of the degenerate star or the
persistence of the X-ray emission.  Alternatively, the difference may
be associated with the angle between the observer and the direction of
source motion.

\item The variability of the lobes and their linear size suggest a
radiative lifetime of less than one hour.  For synchrotron losses in
the emitting plasma, a magnetic field of $\sim 300$ G is needed, which
implies a lobe dominated by magnetic energy.  Alternatively, an
adiabatic expansion of the ultra-relativistic plasma, after generation
at the working surface, is also consistent with the lobe emission
characteristics.

\item The simple pressure balance in the ultra-relativistic plasma on
both sides of the shock at the working surface predicts an advance
speed of the lobe in the range of that observed.  The relatively
constant speed of the lobes over many hours is surprising and no
explanations are offered.

\item The lobe looks like a wagon wheel.  The working surface is at
the hub and the electrons diffuse with 0.57c away rom the
working-surface shock.  The advance of the lobe over the lifetime of
the radiating electrons (30 minutes) also contributes to its size
inthe direction of motion.

\item The slow expansion of the lobes, with a decrease in flux density
and steepening spectrum, may be associated with a decreasing beam flow
and exlarging of the working surface, as well as from adiabatic
expansion of material in the lobes.

\item The \lq scaling\rq~of the Sco X-1 phenomena to that of radio
galaxies suggests that hot-spot variability and life-times may be only
$10^5$ yr in radio galaxy lobes.

\end{itemize}

    These observations show that VLBI resolutions are sufficient to
determine the internal properties of Galactic-jet radio sources.  The
time scales of the evolution vary from less than one hour to a few
days or months in some cases.  Hence, detailed, long-term,
multi-frequency and reasonably continuous observations are required to
follow and understand the emission properties of the sources.  Because
of the possible dominant effect of the magnetic field, linear
polarization imaging would be very useful.

\acknowledgments

The National Radio Astronomy Observatory is a facility of the National
Science Foundation, operated under cooperative agreement by Associated
Universities, Inc.  We thank the European VLBI Network (EVN) and the
Asia-Pacific Telescope (APT) for their support and observation time.
The data were correlated with the VLBA correlator in Socorro and the
Penticton Correlator that is supported by the Canadian Space Agency.
It is a pleasure to thank Dr.\ Jean Swank for granting us RXTE time,
to Dr.\ Tasso Tzioumis for help with the APT scheduling, and to Dr.\
Sean Dougherty for processing the APT data.  We thank Dr.\ Michael
McCollough and Dr.\ Vivek Dhawan for comments on the draft.

\begin{deluxetable}{rllccrr}
\tablewidth{0pt}
\tablecolumns{7}
\tablecaption{The VLBI Observations of Sco X-1}
\tabletypesize{\footnotesize}
\tablehead{
 \colhead {Date}    &
 \colhead {MJD}  &
 \colhead {UT range}  &
 \colhead {Array}  &
 \colhead {Freq}  &
 \colhead {BW} &
 \colhead {Flux Dens.}  \\
 \colhead {} & \colhead {} & \colhead {} & \colhead {} & \colhead
 {(GHz)} & \colhead {(MHz)} & \colhead {(mJy)} \\ } 
\startdata
 19-Aug-1995 & 49948 & 23.7 to 26.7 & VLBA & 4.98 & 64 & 0.4 \\
 16-Mar-1996 & 50158 & 10.1 to 13.0 & VLBA & 4.98 & 64 & 0.6 \\
 14-Sep-1996 & 50340 & 22.1 to 25.0 & VLBA & 4.98 & 64 & 4.5 \\
 03-Aug-1997 & 50663 & 23.9 to 29.3 & VLBA & 4.98 & 64 & 3.5 \\
 21-Aug-1997 & 50681 & 22.8 to 27.8 & VLBA+Y1 & 4.98,1.67 & 128 & 15.0\\ 
 27-Feb-1998 & 50871 & 10.3 to 16.0 & VLBA+Y1 & 4.98,1.67 & 128 &
 22.0 \\ 28-Feb-1998 & 50872 & 10.2 to 16.0 & VLBA+Y1 & 4.98,1.67 &
 128 & 5.0 \\ 29-Aug-1998 & 51054 & 23.8 to 27.0 & VLBA+Y1 & 4.98,1.67
 & 128 & 9.0 \\ \\ 11-Jun-1999 & 51340 & 02.2 to 10.2 &VLBA+Y27 &
 4.98,1.67 & 128 & 20.0 \\ 11-Jun-1999 & 51340 & 10.2 to 18.2 &APT$^1$
 & 4.98 & 64 & 8.0 \\ 11-Jun-1999 & 51340 & 18.2 to 26.2 &EVN+$^2$ &
 4.98 & 128 & 18.0 \\ 12-Jun-1999 & 51341 & 02.2 to 10.2 &VLBA+Y27 &
 4.98,1.67 & 128 & 14.0 \\ 12-Jun-1999 & 51341 & 10.2 to 18.2 &APT$^1$
 & 4.98 & 64 & 12.0 \\ 12-Jun-1999 & 51341 & 18.2 to 26.2 &EVN+$^2$ &
 4.98 & 128 & 14.0 \\ 13-Jun-1999 & 51342 & 02.2 to 10.2 &VLBA+Y27 &
 4.98,1.67 & 128 & 8.0 \\ \enddata \tablecomments{ {\protect \\} 
Y1 = One VLA telescope; Y27 = The entire VLA {\protect \\}$^1$
 The APT consists of Australian Telescope Compact Array, Ceduna,
 Hartebeesthoek Radio Astronomical Observatory, Hobart, Kashima,
 Parkes, Shanghai. {\protect \\} $^2$ The EVN+ consists of Effelsberg,
 Green Bank 140', Hartebeesthoek Radio Astronomy Observatory, Jodrell
 Bank, Medicina, Noto, Onsala, Westerbork.  }
\end{deluxetable}

\clearpage
\begin {figure*}
\vskip 0cm
\hskip 0cm
\epsscale{1.0}
\plotone{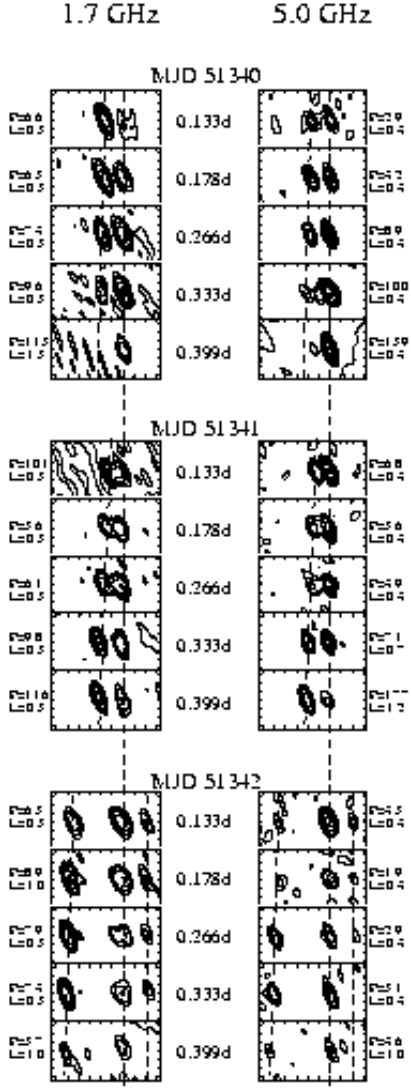}
\vskip 0cm
\caption{
{\bf Images at 1.7 and 5.0 GHz with $10\times 5$ mas
Resolution during 1999 June Observation.} Five images, rotated
$36^\circ$ counter-clockwise in the sky, at 1.7 GHz (left) and 5.0 GHz
(right), are shown for each seven-hour period on three consecutive
days.  The time of each snapshot is shown between the columns.  An
unresolved component will have a position angle of $36^\circ$.  The
minimum contour level (L) and peak flux density (P) in mJy are shown
next to each plot.  The contour levels are $-1,1,2,4,8\ldots$ times
the minimum level.  The tick mark are separated by 10 mas on the
abscissa and 5 mas on the ordinate.  The scale is slightly different
on the two axes.  The dashed lines show the location of the core, NE
and SW components.}

\end{figure*}

\clearpage

\begin {figure*}
\vskip -3cm
\epsscale{2.1}
\plotone{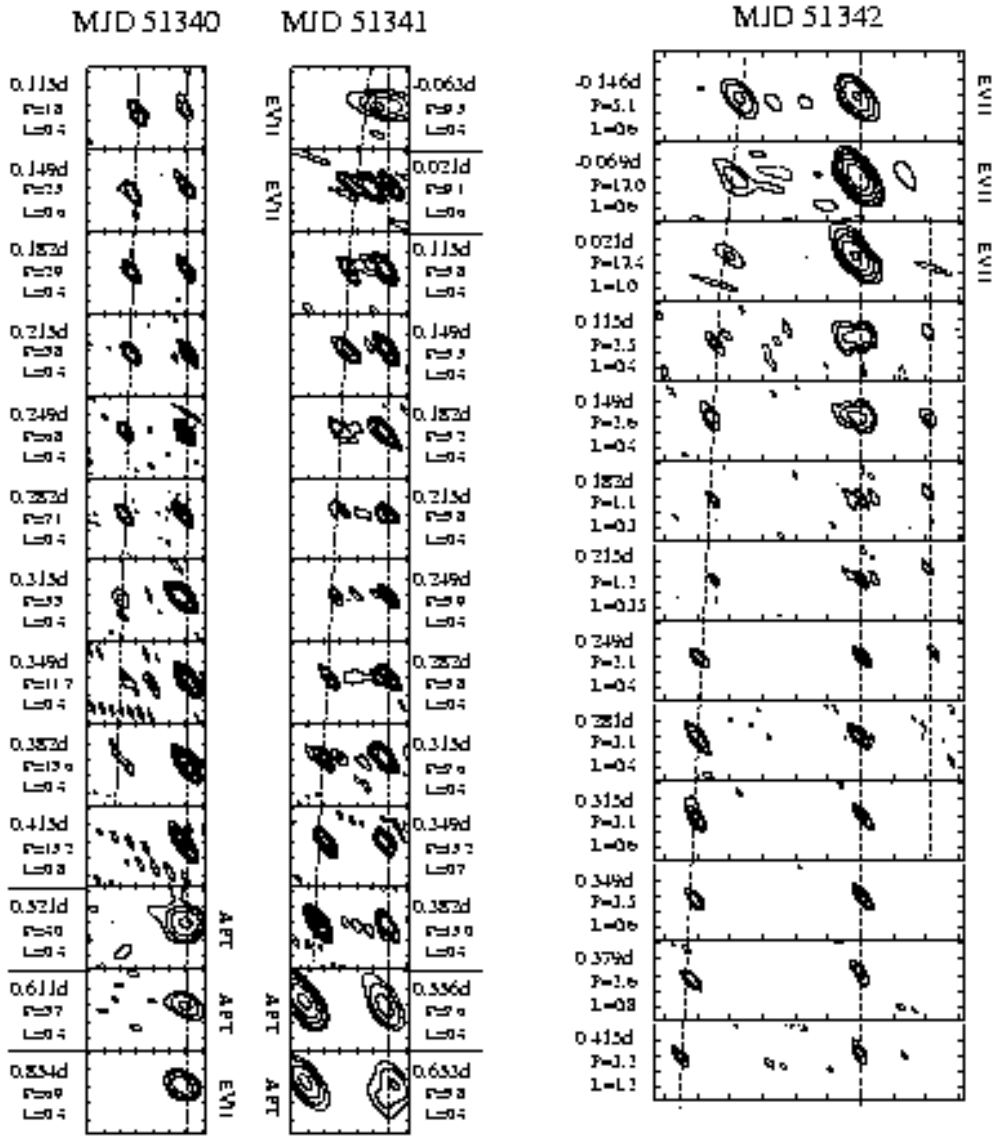}
\caption {
{\bf Images at 5.0 GHz with $4\times 2$ mas Resolution
during 1999 June Observations.}  Thirteen images of Sco X-1, rotated
$36^\circ$ clockwise, are shown for each of the three observing days.
The time of each snapshot, the peak flux density (P) in mJy and the
minimum contour level (L) are given next to each plot.  The contour
levels are -1, 1, 2, 4, 8 $\ldots$ times the minimum level.  The
tick marks are separated by 5 mJy, except for the abscissa on MJD 51342
where they are separated by 10 mJy.  The dashed lines show the
location of the core, the NE component and the SW component on
MJD 51342.  The images from the EVN and APT are labeled.  Others come
from the VLBA.  
}
\end{figure*}

\clearpage

\begin {figure*}
\vskip -4cm
\epsscale{2.2}
\plotone{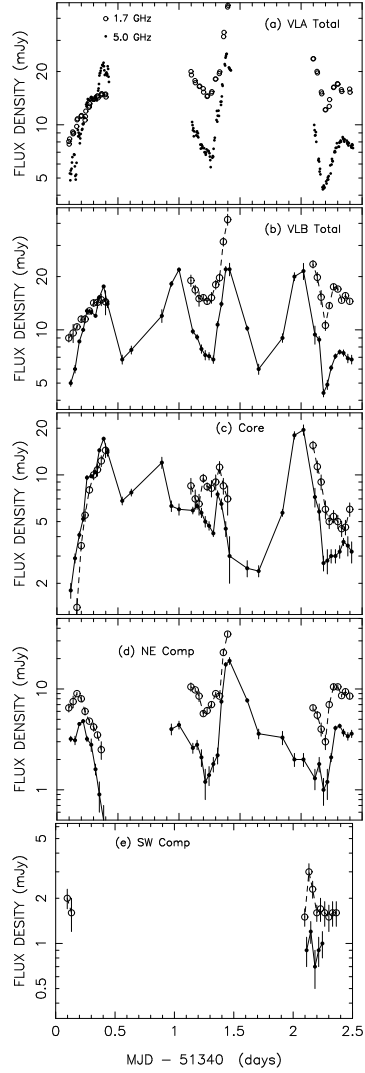}
\vskip -2cm
\caption {
{\bf Flux Density During 1999 June Observations.}  (a) Total Flux
Density from VLA, (b) Total Flux Density from VLBI, (c) Core Flux
Density, (d) NE Component Flux Density, (e) SW Component Flux Density.
The flux densities at 1.7 GHz are shown by the ($\circ$) points, 5.0
GHz by the ($\bullet$). The flux density scale is logarithmic, hence
the flux density difference between frequencies is proportional to the
spectral index.  The one-sigma error bars are indicated.
}
\end{figure*}

\clearpage
\begin {figure*}
\vskip -2cm
\epsscale{1.0}
\plotone{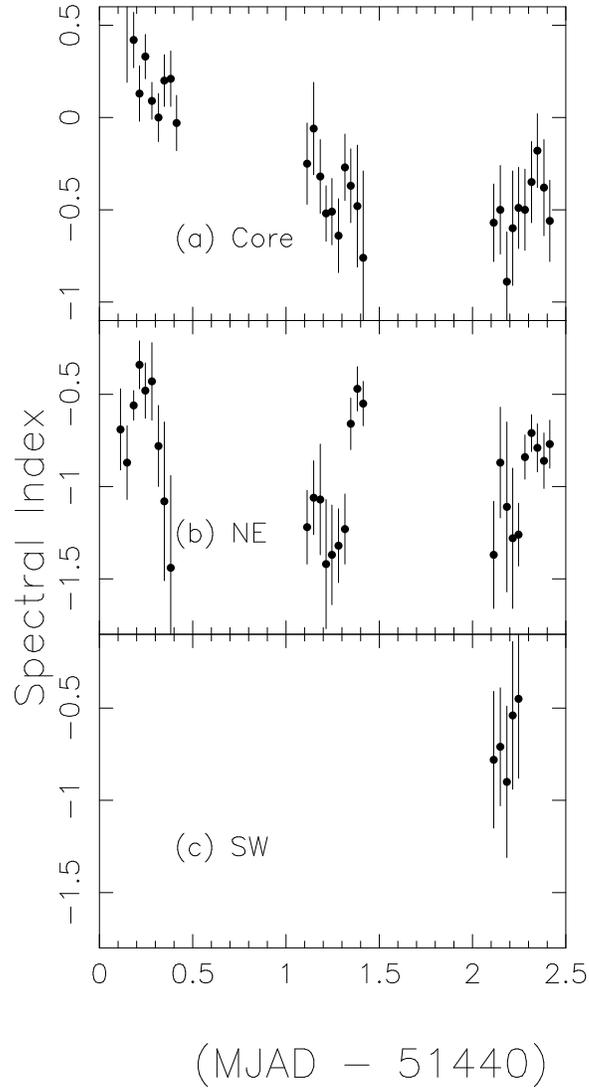}
\vskip 1cm
\caption {
{\bf The Spectral Index During the 1999 June Observations
for the Three Main Components.} (a) Core, (b) NE Component and (c) SW
component.  The spectral indices are calculated only where a detection
at 1.7 GHz and 5.0 GHz was made.  The error bars indicate the
one-sigma error.
}
\end{figure*}

\clearpage

\begin {figure*}
\vskip 0cm 
\epsscale{1.8}
\plotone{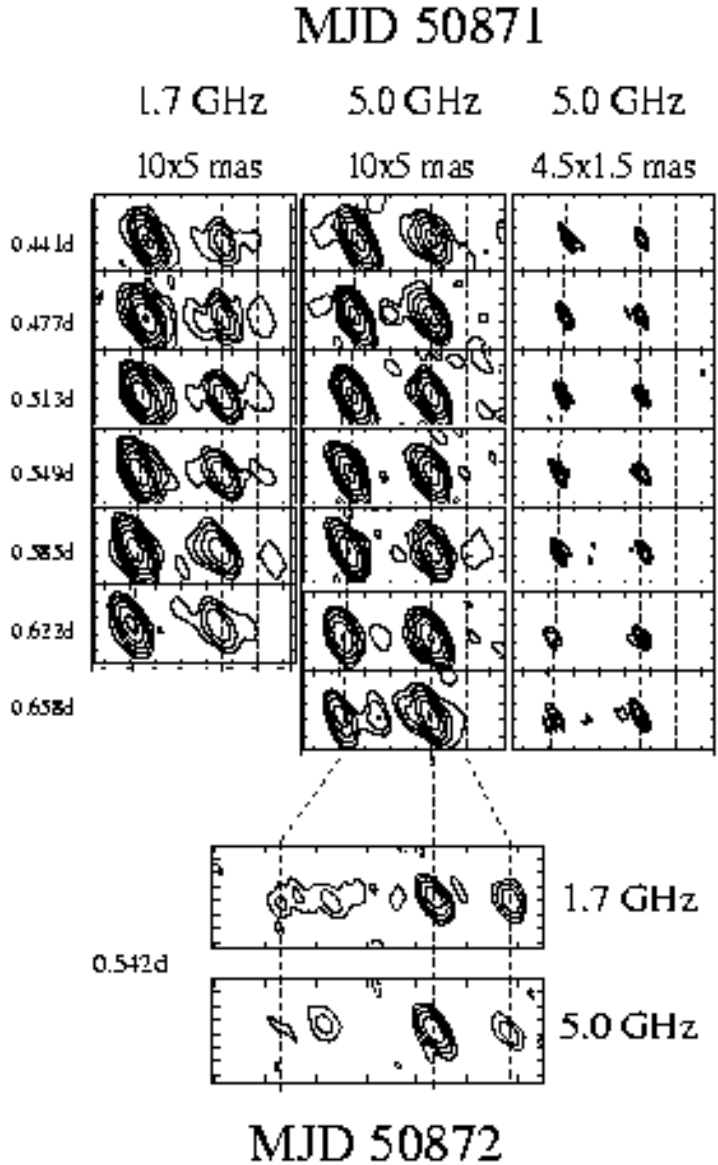}
\caption {
{\bf Images During 1998 February Observations.} The upper
part of the figure shows seven snapshot images on MJD 50871, rotated
by $36^\circ$ on the sky, at the frequency and resolution indicated.
The minimum contour levels are 0.5, 0.4, 0.8 mJy for columns 1, 2 and
3 respectively, with contour levels at $-1,1,2,4,8\ldots$ times the
minimum level.  The tick marks are separated by 10 mas on the abscissa
and 5 mas on the ordinate.  The lower part of the figure shows an
image of Sco X-1 on the following day at 1.7 GHz and 5.0 GHz, each
with a minimum contour level of 0.2 mJy.  The image scale is the same
as for the previous day but the tick marks on the abscissa are 20 mas.
}
\end{figure*}

\clearpage
\begin {figure*}
\vskip -7cm
\epsscale{2.5}
\plotone{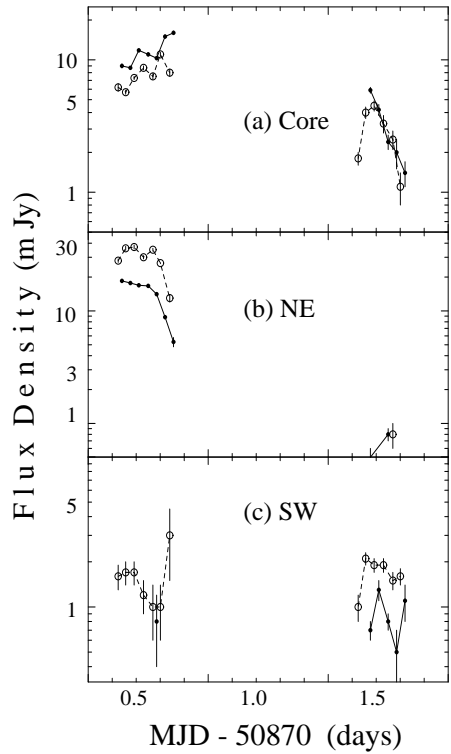}
\vskip -6cm
\caption {
{\bf Flux Density During 1998 February Observations for
the Three Components.} (a) Core, (b) NE Component and (c) SW
component.  Flux densities at 1.7 GHz are shown by the ($\circ$)
points, 5.0 GHz by the ($\bullet$) points.  The flux density scale is
logarithmic and one sigma error bars are indicated.  The points at
each frequency are separated by 50 minutes.
}
\end{figure*}

\clearpage

\begin {figure*}
\vskip -11cm
\epsscale{3.5}
\plotone{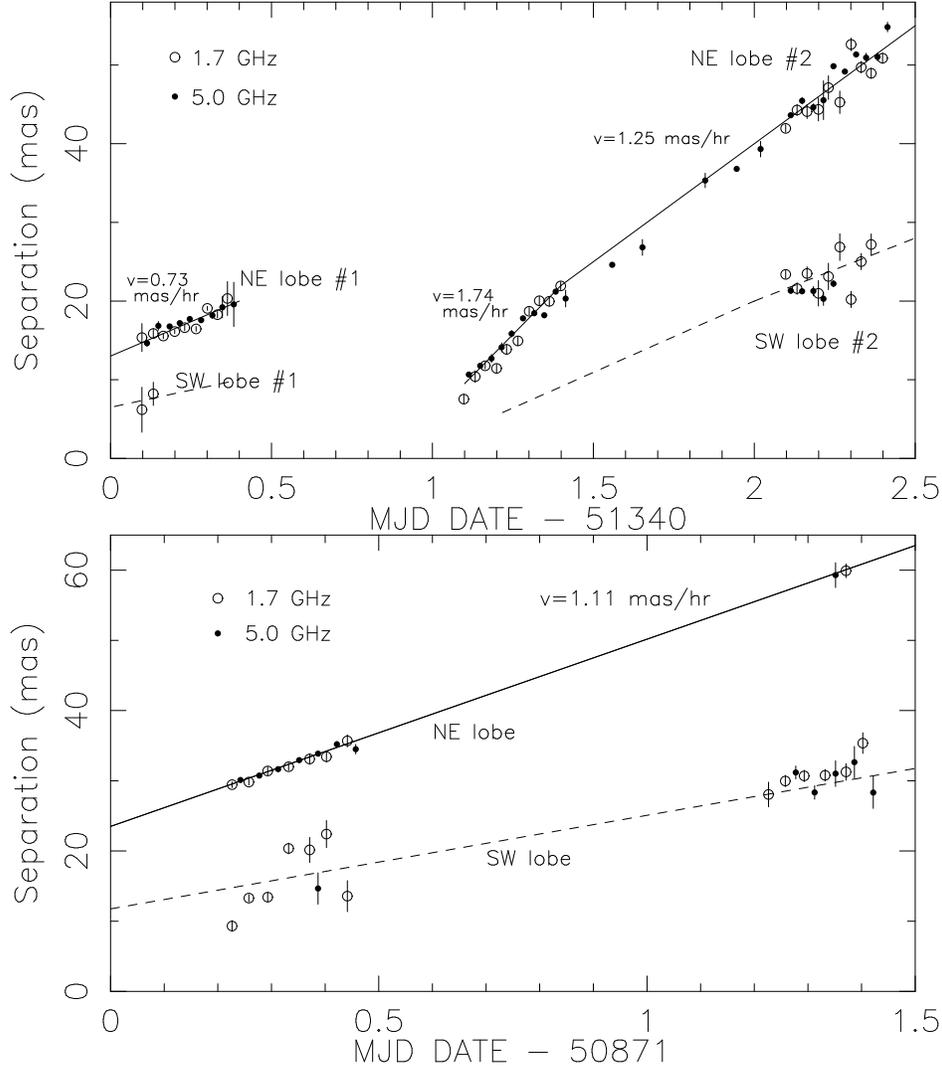}
\vskip -9cm
\caption  {
{\bf The Velocity of the NE and SW Components.} (top) 1999
June observations and (bottom) 1998 February observation: The points
show the separation of the NE and SW components from the core
component during the observations.  The best linear velocity fit for
NE component \#1 and component \#2 in 1999 June and for the NE
component in 1998 February are indicated by the solid lines.  The
dashed lines through the SW components are exactly 50\% of that of the
NE component at the same time.}
\end{figure*}

\clearpage
\begin {figure*}
\vskip -6cm
\epsscale{2.6}
\plotone{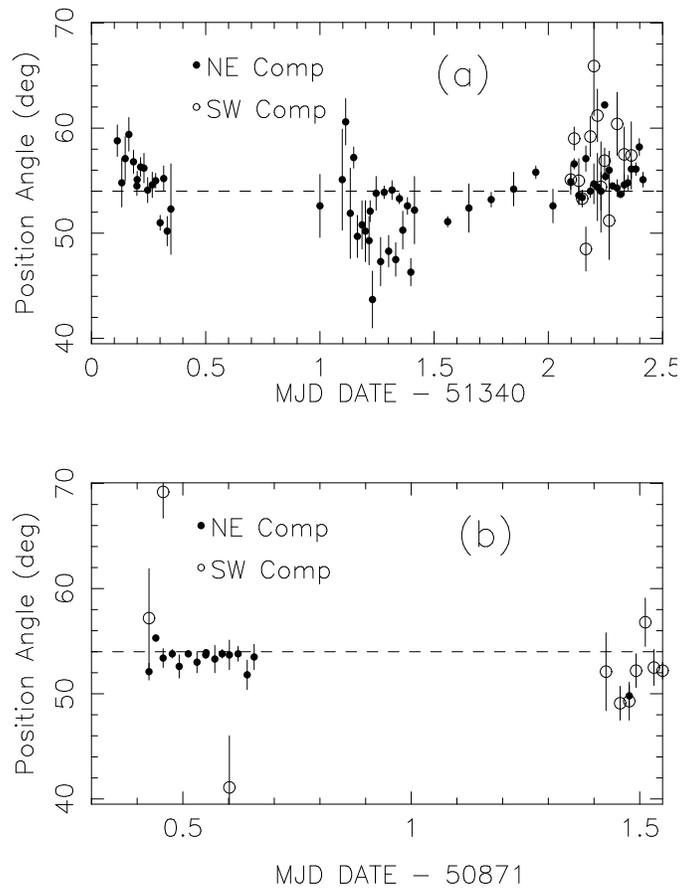}
\vskip -6cm
\caption   {
{\bf The Position Angle of the NE and SW Components.} (a)
The June 1999 observations and (b) the Feb 1998 observations.  The
position angle and error estimate of separation from the core for the
NE components ($\bullet$) and SW components (o) are plotted
separately.  Position angles from both the 1.7 GHz and the 5.0 GHz
observations are plotted.  The dashed-line is at the weighted average
position angle $54^\circ$.}

\end{figure*}

\clearpage

\begin {figure*}
\vskip -8cm
\epsscale{3.1}
\plotone{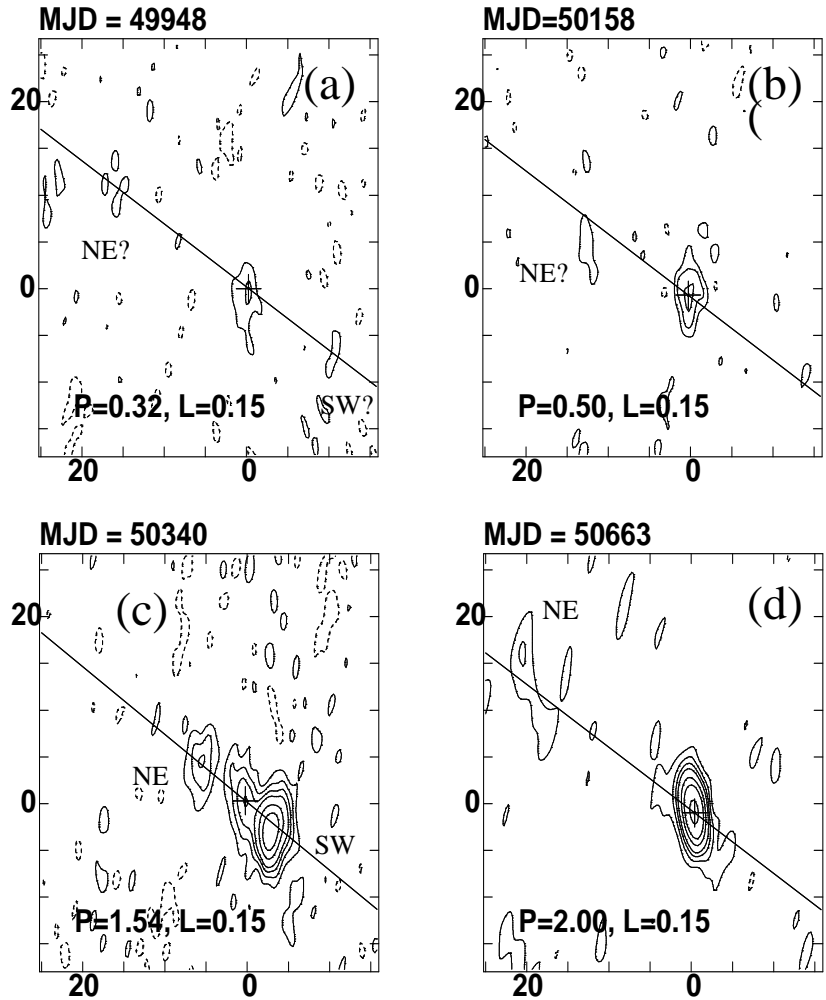}
\vskip -8cm
\caption {
{\bf Images at 5.0 GHz for the first four observation.} (a) MJD
49948, (b) MJD 50158, (c) MJD 50340 and (d) MJD 50663.  The peak flux
density and lowest contour level are shown in each diagram, with
contour levels at -1,1,2,3,4,6,8,12 times the lowest level.  The cross
shows the expected location of the core position as determined from
the radio parallax and proper motion analysis from all observations.
The diagonal line is at position angle $54^\circ$.  Identification of
the NE and SW components (sometimes tentative) are indicated.  The
resolution of each image is $6\times 2$ mas.
}
\end{figure*}

\clearpage
\begin {figure*}
\vskip -3cm
\epsscale{1.3}
\plotone{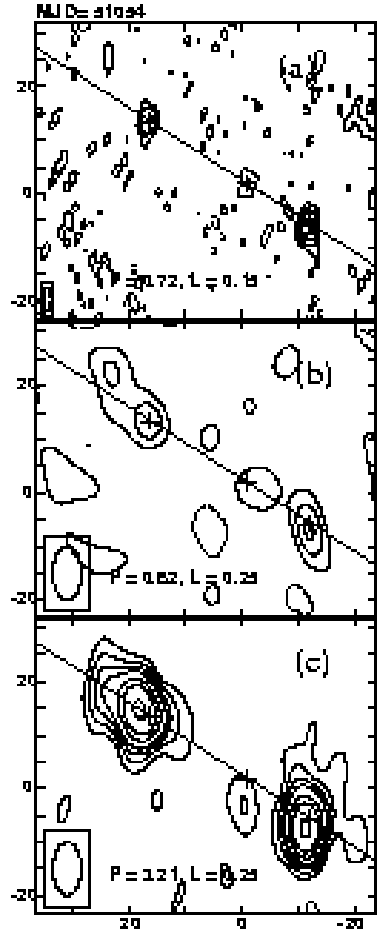}
\vskip -1cm
\caption{
{\bf Images on August 29, 1998.} (a) $4.5\times 1.5$ mas resolution
image at 5.0 GHz, (b) $10\times 5$ mas resolution image at 5.0 GHz and
(c) $10\times 5$ mas resolution image at 1.7 GHz.  The peak flux
density and lowest contour level are shown in each diagram, with
each diagram indicated the resolution.  }
\end{figure*}

\clearpage

\begin {figure*}
\epsscale{2.2}
\vskip -2cm
\plotone{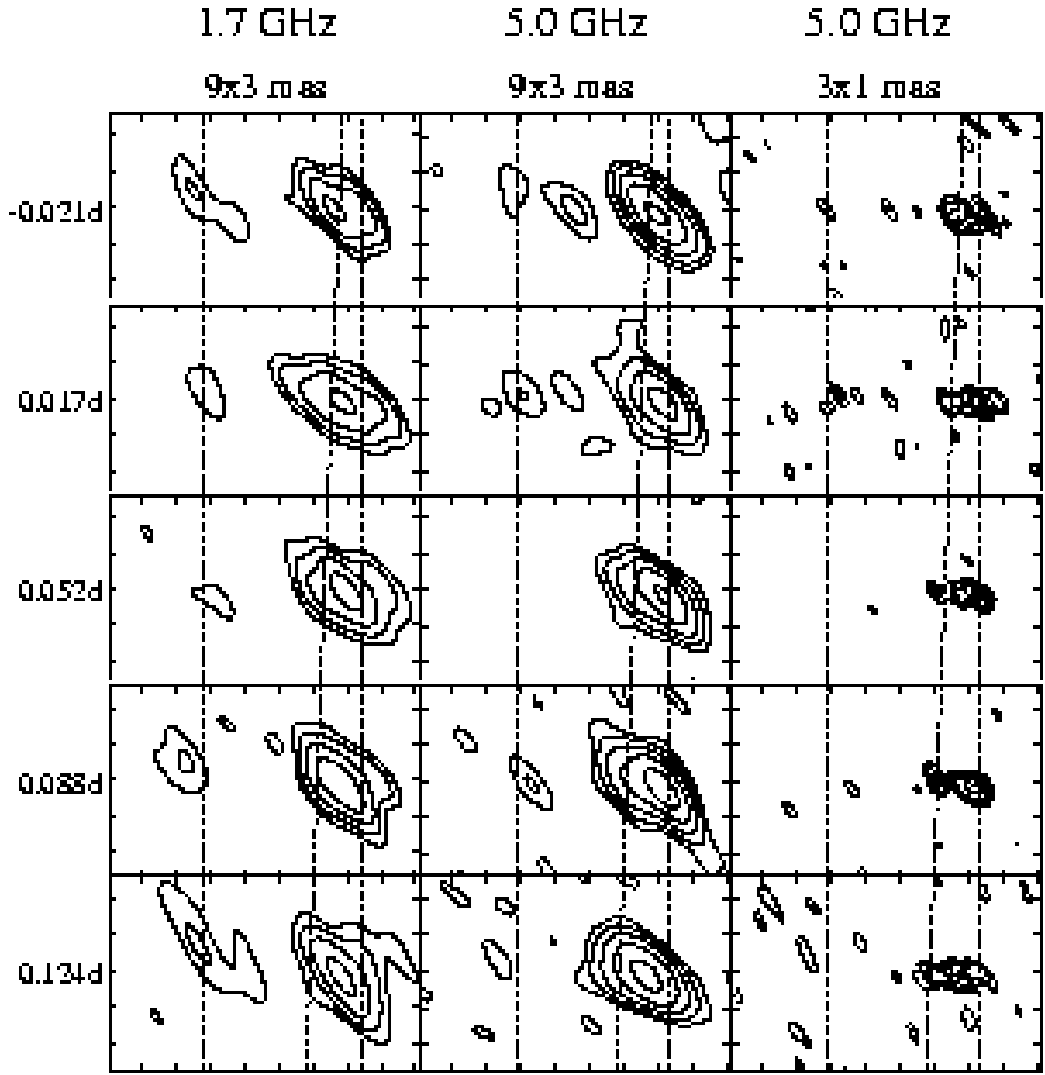}
\vskip -2cm
\caption {
{\bf Images on August 22, 1997.} (left) $9\times 3$ mas
resolution images at 1.7 GHz, (middle) $9\times 3$ mas resolution
images at 5.0 GHz and (right) $3\times 1$ mas resolution images at 5.0
GHz.  All images have been rotated $36^\circ$ counter-clockwise.  The
contour levels are $-1,1,2,4,8\ldots$ times the minimum contour level
of 0.5, 0.4, 0.7 mJy from left to right respectively.  The tick mark
separation is 5 mas in each coordinate.  The vertical dashed line on
the left of each frame is the core position.  The slightly
non-vertical dashed line indicated the approximate location of the NE
component.  The dashed line to the left in each frame is the location
of a possible faint relic NE component.
}
\end{figure*}

\clearpage
\begin {figure*}
\vskip -3cm
\epsscale{2.2}
\vskip 2cm
\plotone{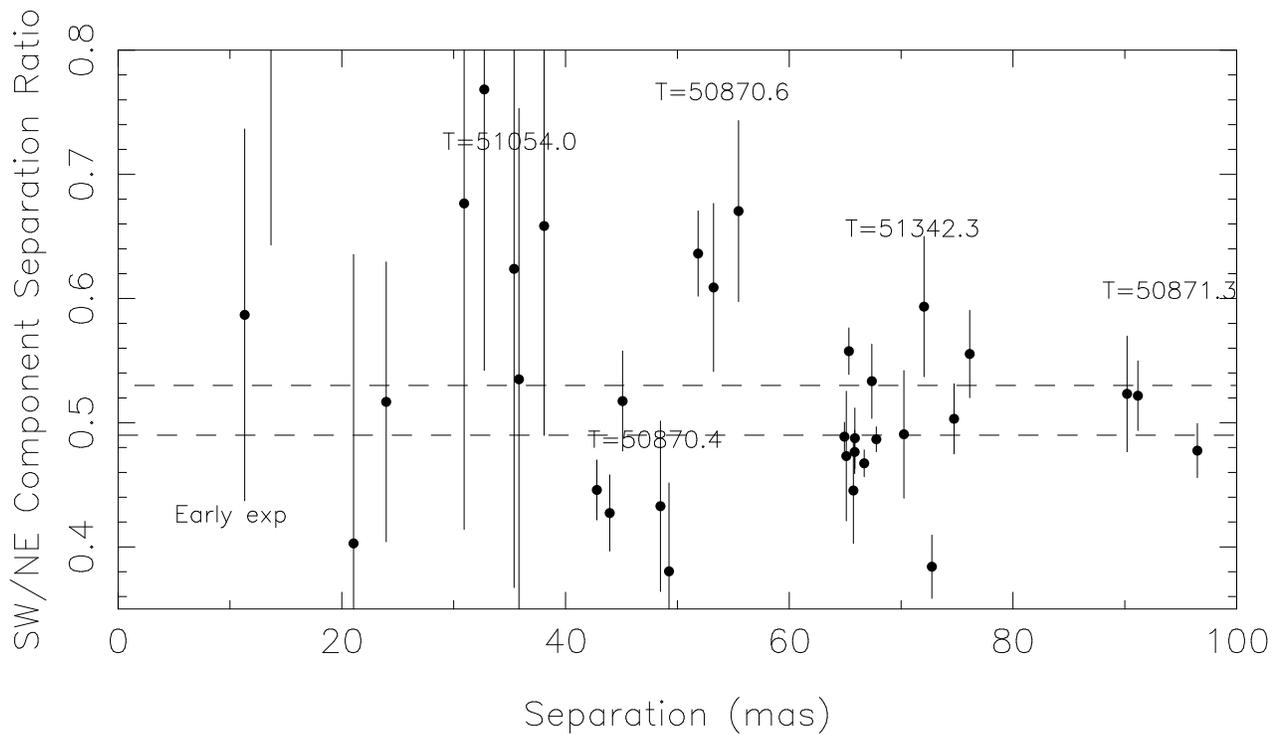}
\caption {
{\bf The Distance Ratio for the SW and NE Components.} The
ordinate is the ratio of the core-SW component separation divided by
the core-NE component separation.  The abscissa is the total
separation between the NE and SW components.  The observation date
(MJD=T) is indicated for the clumps of points, and the error bars
indicate the estimated error.  The ratio $0.51\pm 0.02$ is indicated
by the dashed-lines.  This value is the mean value and error estimate,
heavily weighted by the 1998 February and 1999 June Observations.
}
\end{figure*}

\clearpage

\begin {figure*}
\vskip 0cm
\epsscale{1.6}
\plotone{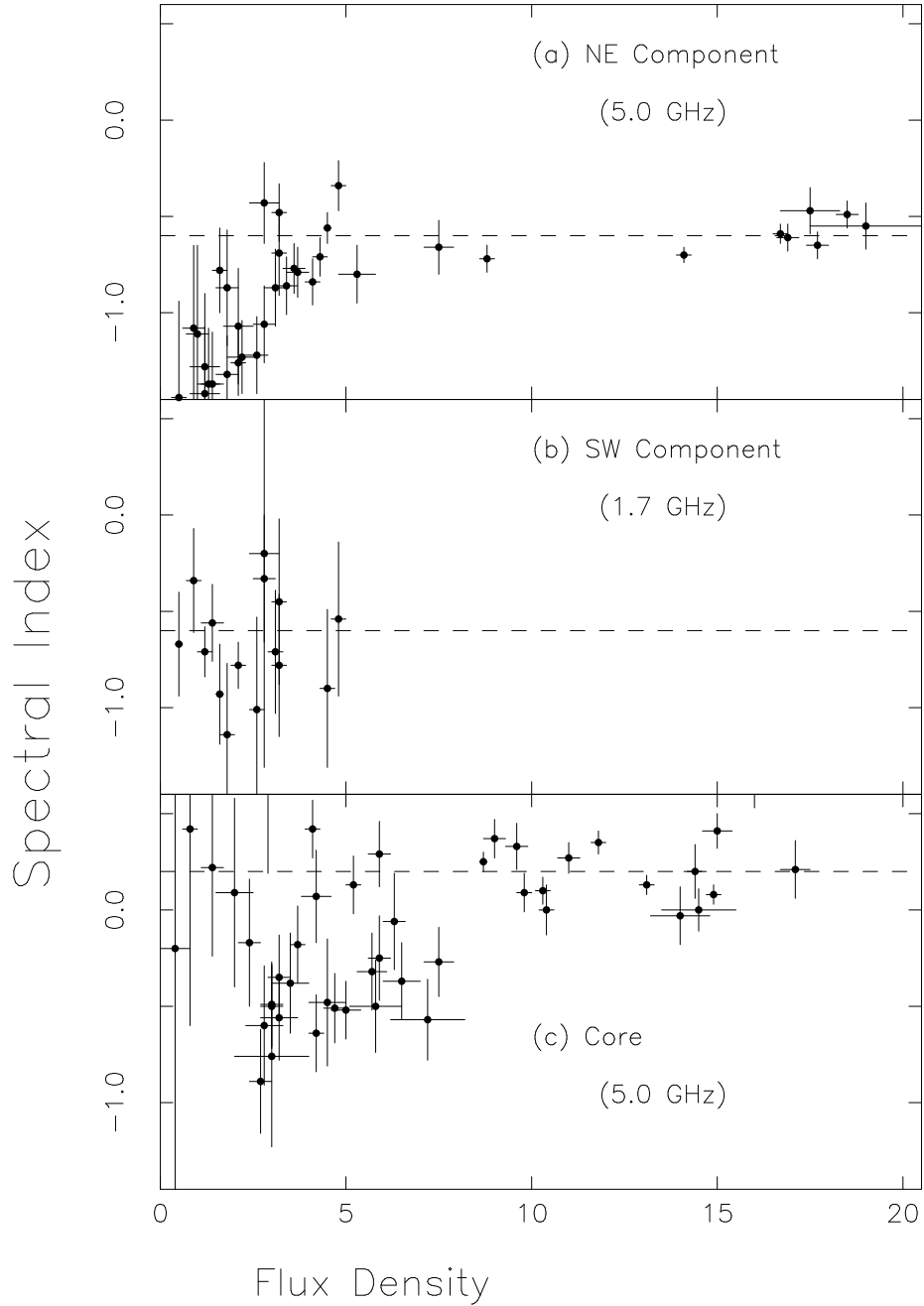}
\caption {
{\bf The Spectral Index vs Flux Density for the
Components.} (a) The NE component at 5.0 GHz, (b) The SW component at
1.7 GHz and (c) The Core at 5.0 GHz.  The spectral index $\alpha$
($S\propto \nu^\alpha$) is plotted versus the flux density (S) of each
component.  Most of these data are from the 1999 June and 1998
February observations.  The dashed line is at $\alpha=-0.6$ for (a)
and (b), and $\alpha=+0.2$ for (c).  The error bars show the estimated
spectral index and flux density errors for each point. The data are
shown at 1.7 GHz for the SW component since it was detected more often
at this frequency.
}
\end{figure*}

\clearpage
\begin {figure*}
\vskip -2cm
\epsscale{2.1}
\plotone{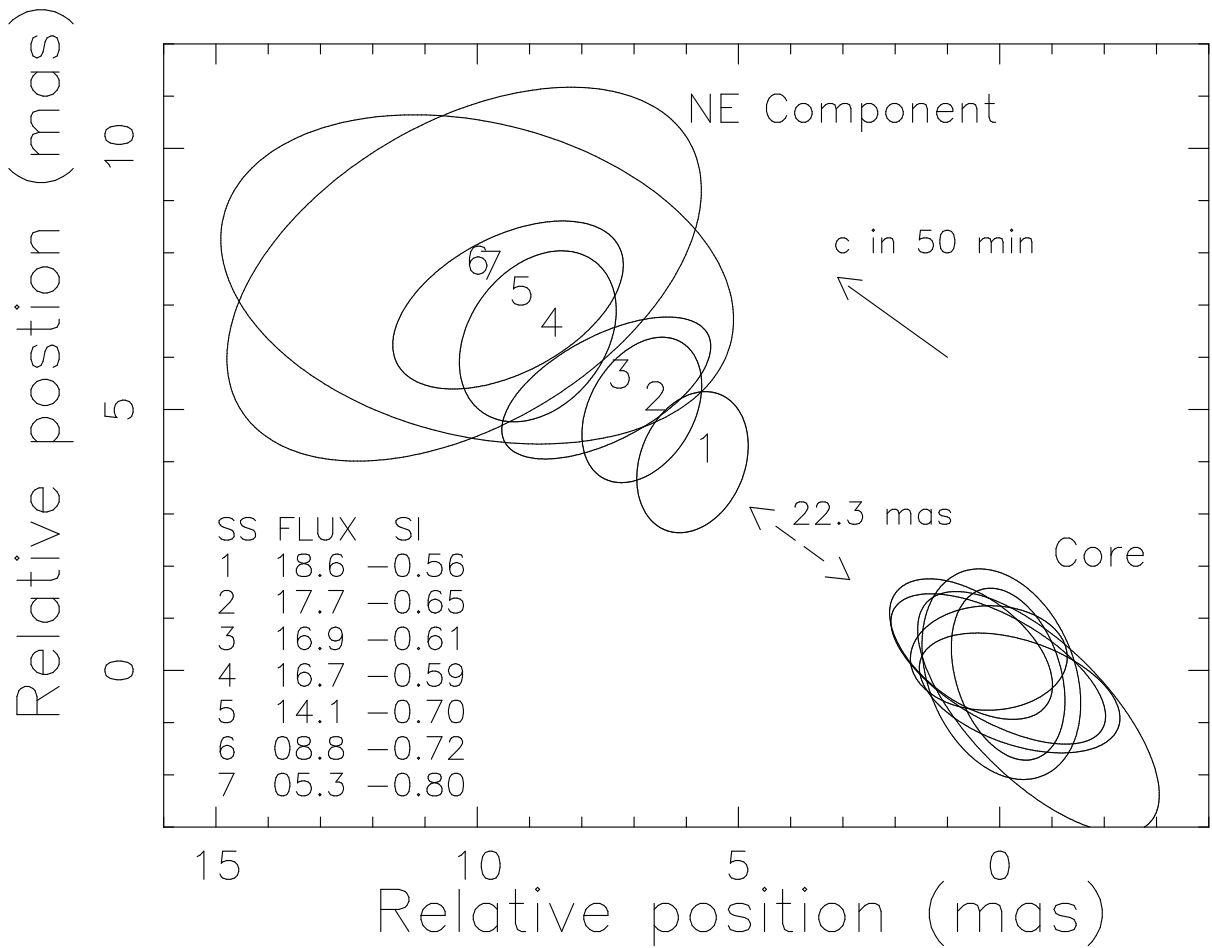}
\vskip 0cm
\caption {
{\bf  The Expansion of the NE component on MJD 50871.} The
position and full-width half-intensity angular size of the NE
component and core component on MJD 50871 are shown by the ellipses.
The components have been moved closer together by 22.3 mas in position
angle $54^\circ$ to fit on the plot.  The numbers at the center of the
NE component ellipses indicate the snapshot number (SS) which are
separated in time by 50-minutes.  No snapshot number is given for the
core component since it is more stationary.  The arrow labeled 'c in
50 min' shows the distance that a light signal will travel in the
plane of the sky in 50 minutes.  For each snapshot (SS), the total
flux density (FLUX) and spectral index (SI) is given for the NE
component.
}
\end{figure*}

\clearpage

\begin {figure*}
\vskip -3cm
\epsscale{1.8}
\plotone{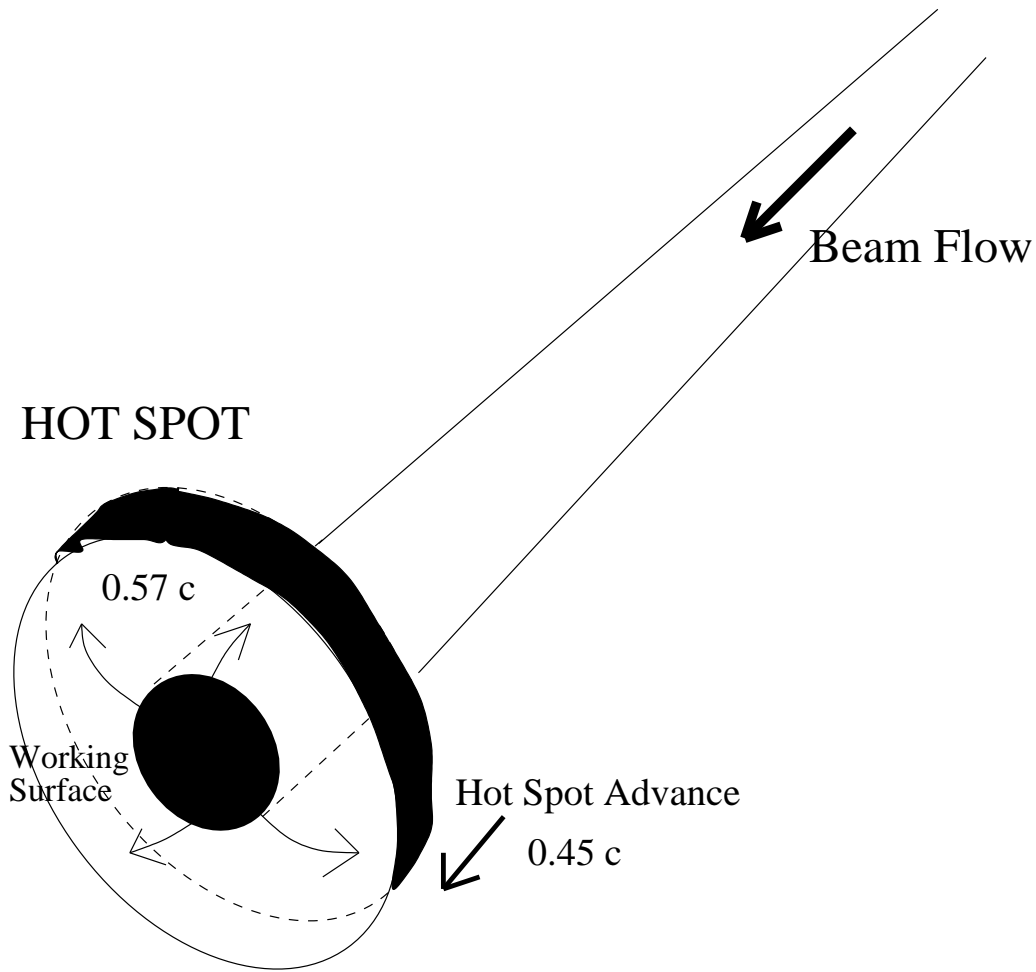}
\vskip 2cm
\caption {
{\bf Schematic of Lobe in Sco X-1.} A beam of energy from
the binary region of Sco X-1 interacts with the ambient material to
form a working surface composed of an ultra-relativistic plasma.
Radiating electrons diffuse with $v=0.57$c along the radial magnetic
field lines to form the lobe.  The working surface is also moving at
$\approx 0.45$c at $45^\circ$ to the line of sight.  The minimum
diameter of the lobe is 1.5 mas ($3\times 10^{8}$ km) and 0.6 mas
($1.3\times 10^{8}$ km) perpendicular and parallel, respectively, to
the lobe motion.
}
\end{figure*}
\clearpage

\begin {figure*}
\vskip -9cm
\epsscale{3.0}
\plotone{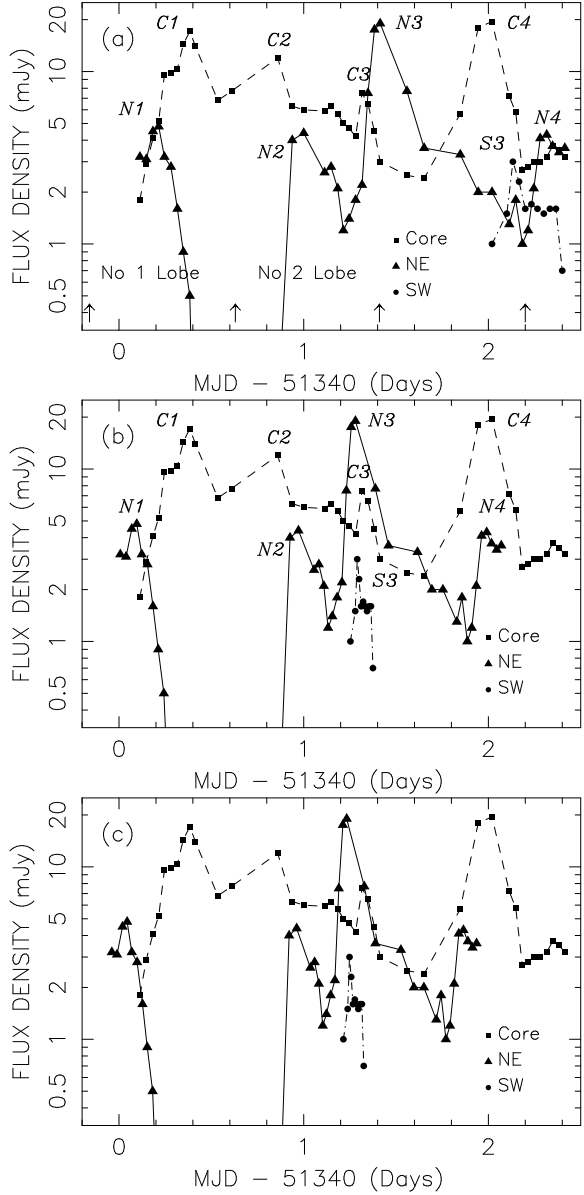}
\vskip -8cm
\caption {
{\bf The Correlation of Variations in the Core and Lobes
in 1999 June.} (a) The observed flux densities of the core and both
lobes, (b) the flux densities with the NE and SW lobes corrected for a
beam speed of $\beta_j=1.0$ and (c) the flux densities with the NE and
SW lobes corrected for a beam speed of $\beta_j=0.9$.  The flux
density scale is logarithmic.  The major lobe flares for the core are
C1, C2, C3, C4; for the NE lobe are N1, N2, N3, N4; for the SW lobe,
the only detected flare, is indicated by S3.  The arrows on the
abscissa of (a) indicated the time of minimum light for the binary
system.
}
\end{figure*}

\clearpage

\end{document}